\documentclass[%
 reprint,
showpacs,
amsmath,amssymb,
aps,
prb,
]{revtex4-1}

\usepackage{graphicx}
\usepackage{dcolumn}
\usepackage{bm}

\def\be{\begin{equation}}
\def\ee{\end{equation}}
\def\bea{\begin{eqnarray}}
\def\eea{\end{eqnarray}}
\def\kv{{\bf k}}

\def\no{\nonumber}
\def\ra{\rangle}
\def\la{\langle}
\def\s{\sigma}
\def\t{\tau}
\def\ve{\varepsilon}

\usepackage{color}

\newcommand{\dg}[1]{{#1}^\dagger}
\newcommand{\bra}[1]{\left\langle{#1}\right|}
\newcommand{\ket}[1]{\left|{#1}\right\rangle}

\begin{document}

\title{Inhomogeneous hardcore bosonic mixture with checkerboard supersolid phase:\\ Quantum and thermal phase diagram}

\author{F. Heydarinasab}
\email[]{fheydari@iasbs.ac.ir}
\affiliation{Department of Physics, Institute for Advanced Studies in Basic Sciences (IASBS), Zanjan 45137-66731, Iran}
\author{J. Abouie}
\email[]{jahan@iasbs.ac.ir}
\affiliation{Department of Physics, Institute for Advanced Studies in Basic Sciences (IASBS), Zanjan 45137-66731, Iran}

\begin{abstract}
We introduce an inhomogeneous bosonic mixture composed of two
kinds of hardcore and semi-hardcore boson with different nilpotency conditions and demonstrate
that in contrast with the standard hardcore Bose Hubbard model, our
bosonic mixture with nearest and next nearest neighbor interactions
on a square lattice develops the checkerboard supersolid phase
characterized by the simultaneous superfluid and checkerboard solid
orders. Our bosonic mixture is created from a two-orbital Bose-Hubbard model including two kinds of bosons: a single orbital boson and a two-orbital boson.
By mapping the bosonic mixture to an anisotropic
inhomogeneous spin model in the presence of a magnetic field, we
study the ground state phase diagram of the model by means of
cluster mean field theory and linear spin wave theory and show that
various phases such as solid, superfluid, supersolid and Mott
insulator appear in the phase diagram of the mixture. Competition
between the interactions and magnetic field causes the mixture to
undergo different kinds of first and second order phase transitions.
By studying the behavior of the spin wave excitations we find the
reasons of all first and second order phase transitions. We also
obtain the temperature phase diagram of the system using cluster
mean field theory. We show that the checkerboard supersolid phase
persists at finite temperature comparable with the interaction
energies of bosons.
\end{abstract}

\date{\today}


\maketitle


\section{Introduction}\label{sec:intro}
Supersolids are characterized by the simultaneous presence of a
nontrivial crystalline solid order and superfluid phase order in the
context of quantum lattice gas models\cite{Matsuda701, Liu1973}.
Discussing the possibility of supersolidity, has attracted renewed
interest in connection with ultracold Bose gases in optical
lattices\cite{Griesmaier05, Deiglmayr08, Ni231, Lu11, Aikawa12,
Aikawa10}. The precise controllability of optical lattice systems
has motivated theoretical explorations of supersolid phase in
various systems, such as one dimensional
chains\cite{PhysRevLett.97.087209, sengupta2007spin,
PhysRevA.79.011602, PhysRevB.80.174519}, two dimensional
square\cite{PhysRevLett.88.170406, 0295-5075-72-2-162,
PhysRevLett.94.207202, PhysRevLett.95.033003, PhysRevLett.98.260405,
PhysRevLett.103.225301, PhysRevLett.104.125301, PhysRevA.82.013645,
Mila01062008, PhysRevB.86.054516, ng2008supersolid},
honeycomb\cite{PhysRevB.75.174301, PhysRevB.75.214509},
triangular\cite{PhysRevLett.95.127205, PhysRevLett.95.127206,
PhysRevLett.95.127207, PhysRevLett.95.237204, PhysRevB.76.144420,
PhysRevLett.100.147204, PhysRevLett.104.125302, PhysRevB.84.054510,
PhysRevB.84.174515, PhysRevA.85.021601} and
kagome\cite{PhysRevLett.97.147202} lattice structures,
two-dimensional spin-$1/2$ dimer model with an anisotropic
intra-plane antiferromagnetic coupling\cite{PhysRevLett.97.127204},
bilayer systems of dipolar lattice
bosons\cite{PhysRevLett.103.035304} and three dimensional cubic
lattice\cite{PhysRevLett.98.260405, PhysRevB.79.094503,
PhysRevB.84.054512, PhysRevLett.108.185302}. These extensive studies
show that no supersolid phases can exist in the ground state phase
diagram of the hardcore Bose Hubbard model with nearest neighbor
interaction for bipartite lattices\cite{PhysRevLett.84.1599,
PhysRevB.75.174301, PhysRevB.75.214509, ng2008supersolid, schmid2002finite, hebert2001quantum, PhysRevB.86.054516}. In these systems, due to
the formation of antiphase walls between ordered
domains\cite{PhysRevB.68.014506}, supersolid states are unstable
towards phase separation\cite{PhysRevLett.94.207202}. In order to
have stable supersolid phases, one has to modify the model by
introducing repulsive dipole-dipole
interaction\cite{PhysRevLett.104.125301, doi:10.1143/JPSJ.80.113001,
PhysRevB.86.054516} where has the role of increasing the energy cost
of domain wall formations. Adding next nearest neighbor
interaction\cite{PhysRevB.51.8467, PhysRevLett.84.1599, ng2008supersolid, hebert2001quantum, dang2008vacancy}, correlated
hoppings \cite{PhysRevLett.100.090401, Mila01062008}, or treating
soft core bosons\cite{0295-5075-72-2-162, PhysRevLett.94.207202},
two-component Bose-Fermi\cite{PhysRevLett.91.130404,
PhysRevLett.100.100401, orth2009supersolidity} and
Bose-Bose\cite{PhysRevB.80.245109, keilmann2009dynamical} mixtures,
and three component Bose-Bose-Fermi mixture\cite{yan2013supersolid}
result also stable supersolids.

In this paper we introduce a different inhomogeneous bosonic model (IBM) which is composed of two kinds of hardcore and semi-hardcore boson, $a$ and $b$,
with {\it different nilpotency conditions}: $(\dg a_i)^2=0$ for $a$
and  $(\dg b_i)^3=0$ for $b$ bosons, and show that the model on a
square lattice with nearest neighbor (NN) and next nearest neighbor
(NNN) interactions is an appropriate ground for searching various
supersolid orders. The nilpotency condition for $b$ bosons signifies
that one can put up two $b$ particles on each lattice site.
Our IBM is created from a Bose-Hubbard model including two kinds of
bosons: a single orbital boson and a two-orbital boson.
By mapping the IBM to an anisotropic
inhomogeneous spin-(1,1/2) model in the presence of a magnetic field, we
study the ground state phase diagram of the model by means of
cluster mean field (CMF) theory and linear spin wave (LSW) theory and show that
various phases such as solid, superfluid, supersolid and Mott insulator appear in the phase diagram of the mixture.
We demonstrate that in contrast with the standard hardcore Bose Hubbard model
in which long range hopping terms are required for the superfluidity\cite{PhysRevLett.100.090401, Mila01062008},
or long range dipole-dipole interactions are necessary to suppress quantum fluctuations for the stability of checkerboard supersolid (CSS) order
on the square lattice\cite{PhysRevB.86.054516, PhysRevLett.104.125301, PhysRevLett.103.225301, doi:10.1143/JPSJ.80.113001},
our IBM possesses an stable CSS phase even in the absence of long range interaction and long range hopping terms.
This stability is attributed to the difference in the nilpotency conditions of $a$ and $b$ bosons.
The small amount of spin wave fluctuations also show the stability of the CSS phase.
Making use of LSW theory and obtaining the excitation spectra of the IBM,
besides the strength of quantum fluctuations around the mean field ground states we find the boundaries of the stability of the mean field phases.

In this paper we also study the effects of temperature on the phase
diagram of the system. We obtain the temperature phase diagram of
the mixture and show that in the presence of temperature various phases emerge in the phase diagram. Our results show that the CSS order
can persist even at finite temperatures comparable with the
interaction energies.

This paper is organized as follows. In section \ref{sec:model} we
introduce our IBM and map the model onto a mixed spin model by making
use of hardcore boson-spin transformations. In section \ref{sec:CMF}
we give a brief review on the CMF theory and generalize the theory to the mixed spin model. By computing the diagonal and off diagonal order
parameters we present the CMF ground state phase diagram of the
model in section \ref{sec:GPD}. In order to investigate the
stability of these phases against quantum fluctuations we compute the
order parameters within CMF theory with larger clusters. The
strengths of quantum fluctuations for each phase are also obtained
by means of LSW theory in section \ref{sec:LSW}. In this section we
investigate the behavior of the spin wave dispersions at phase
transitions to figure out the reason of all first and second order
phase transitions in the phase diagram of the IBM. In the
second part of the paper, in section \ref{sec:TPD}, we obtain the thermal
phase diagram of the model and show that the CSS order survives, at finite temperatures. Finally, we summarize our
results and give the concluding remarks in section
\ref{sec:summary}.


\section {Inhomogeneous bosonic model}\label{sec:model}
Let us consider the two kinds of hardcore and semi-hardcore boson $a$ and $b$ which interact via the Hamiltonian:
\begin{eqnarray}
\no H_B&=& -t\sum_{\la i,j\ra}(\dg a_i b_j+\dg b_j a_i)+U\sum_i n^b_i n^b_i\\
\no &+& V_1\sum_{\la i,j\ra}n^a_i n^b_j
+V_2\sum_{\la\la i,j\ra\ra}(n^a_i n^a_j+n^b_i n^b_j)\\
\label{BH1} &-& \sum_i (\mu^a n^a_i + \mu^b n^b_i),
\end{eqnarray}
where $\dg a_i(a_i)$ and $\dg b_j(b_j)$ are respectively the
creation(annihilation) operators of $a$ and $b$ particles at sites $i$ and $j$, on a two dimensional (2D) bipartite square lattice. The first term represents a hopping between two nearest neighbor sites $\la i,j\ra$ where $i$ and $j$ are the lattice points in subsystems I and II, respectively. $U$ is local Coulomb attraction energy ($U<0$) between $b$ bosons occupying the same site, $V_1$ is the interaction energy between $a$ and $b$ bosons, $V_2$ denotes the interaction between two $a$ or two $b$ bosons, and $\mu^a$ and $\mu^b$ are chemical potentials. $\la \dots\ra$ and $\la\la \dots\ra\ra$ indicate the summations over nearest and next nearest neighbors on the square lattice, respectively.

The $a$ particles are canonical hardcore bosons and satisfy the
canonical commutation relations. The number of these bosons at site
$i$ is $n^a_i=\dg a_i a_i$, and the nilpotency condition for them is
$(a_i^\dagger)^2=0$. The $b$ particles are however semi-hardcore
bosons and satisfy the nilpotency condition $(\dg b_i)^3=0$, which
signifies that one can put up two $b$ particles on each lattice
site. This uncommon nilpotency condition leads to the following
non-canonical algebra (see appendix \ref{appendixb}, for the detailed
calculation):
\begin{eqnarray}
\no &&[b_i,b_j]=[\dg b_i, \dg b_j]=0,\\
&&[b_i, \dg b_j]=\delta_{ij}(1-n^b_i),~~~~ [n^b_i, \dg b_j]=\delta_{ij}\dg b_j,
\label{commutation-b}
\end{eqnarray}
where $n^b_j(\neq \dg b_jb_j)$ is the number of $b$ bosons which possesses the
relation $\dg {(n^b_i)}=n^b_i$.

Since the number operator $n^b$ in not equal to $\dg b b$, the Hamiltonian in Eq. (\ref{BH1}) does not have the standard form
of a Bose Hubbard Hamiltonian. But, as we will show in appendix \ref{appendixa}, this Hamiltonian is created from an standard two-orbital bosonic Hubbard
model (see Eq. (\ref{2FBH}) in appendix \ref{appendixa}) by reducing effective
number of degrees of freedom. The three-body constraint of the
semi-hardcore bosons $b$, and consequently their non-canonical
statistics algebra arise inevitably from the transformations in
Eq. (\ref{defSA}) which are employed for mapping the two-orbital
Hamiltonian in Eq. (\ref{2FBH}) to the one in Eq. (\ref{BH1}). At the first glance it may seem that the non-canonical statistics of $b$ particles makes the model complicated, but as we will show in next sections,
using a simple boson-spin transformation, the Hamiltonian (\ref{BH1}) maps to an standard spin Hamiltonian with rich phase diagrams.

It is worth to mention that two independent physical properties are
responsible for the quantum statistics of particles. The first one
is exchange or permutation statistics which concerns braiding of
particles and the second one is exclusion statistics which concerns
number of particles allowed to occupy the same
site\cite{Batista2004}. It should be noted that, although the
commutation relations in Eq. (\ref{commutation-b}) have fractional
exclusion statistics, they obey canonical exchange statistics, and
should not be confused with the anyonic particles with fractional
exchange statistics\cite{aidelsburger2013realization,
miyake2013realizing, greschner2013ultracold, yao2012topological,
yao2013realizing,   PhysRevLett.101.260501}. The anyonic algebra can
be created and manipulated by using the so-called conditional
hopping terms\cite{keilmann2010statistically,
PhysRevLett.117.205303, PhysRevA.94.013611, atala2014observation} which is not the case in
our paper.

Fractional exchange statistics in bosonic systems causes the system
to experience different new phases which are not seen in the
standard system with canonical bosons. For example, the one
dimensional optical lattice of semi hardcore bosons with the
constraint $(b^{\dagger}_i)^3=0$ and fractional statistics, proposed
by Greschner, \textit{et. al.}\cite{greschner2015anyon}, shows a
novel two-component superfluid of holon and doublon dimers,
characterized by a large but finite compressibility and a
multipeaked momentum distribution, which is not seen in the one
dimensional canonical model\cite{Daley2009}. Moreover, including
such an statistics in the dipolar system with hardcore bosons
results in an striped supersolid phase\cite{yao2012topological}. In
Eq. (\ref{BH1}) we have introduced an inhomogeneous system of
hardcore and semi-hardcore bosons which could be realized in a
two-orbital bosonic system. The non-canonical statistics of the
semi-hardcore bosons causes the model to be mapped to a mixed spin
model which possesses the CSS phase, in addition to the superfluid, Mott insulating and various solid phases.
\begin{figure}[t]
\centerline{\includegraphics[width=80mm]{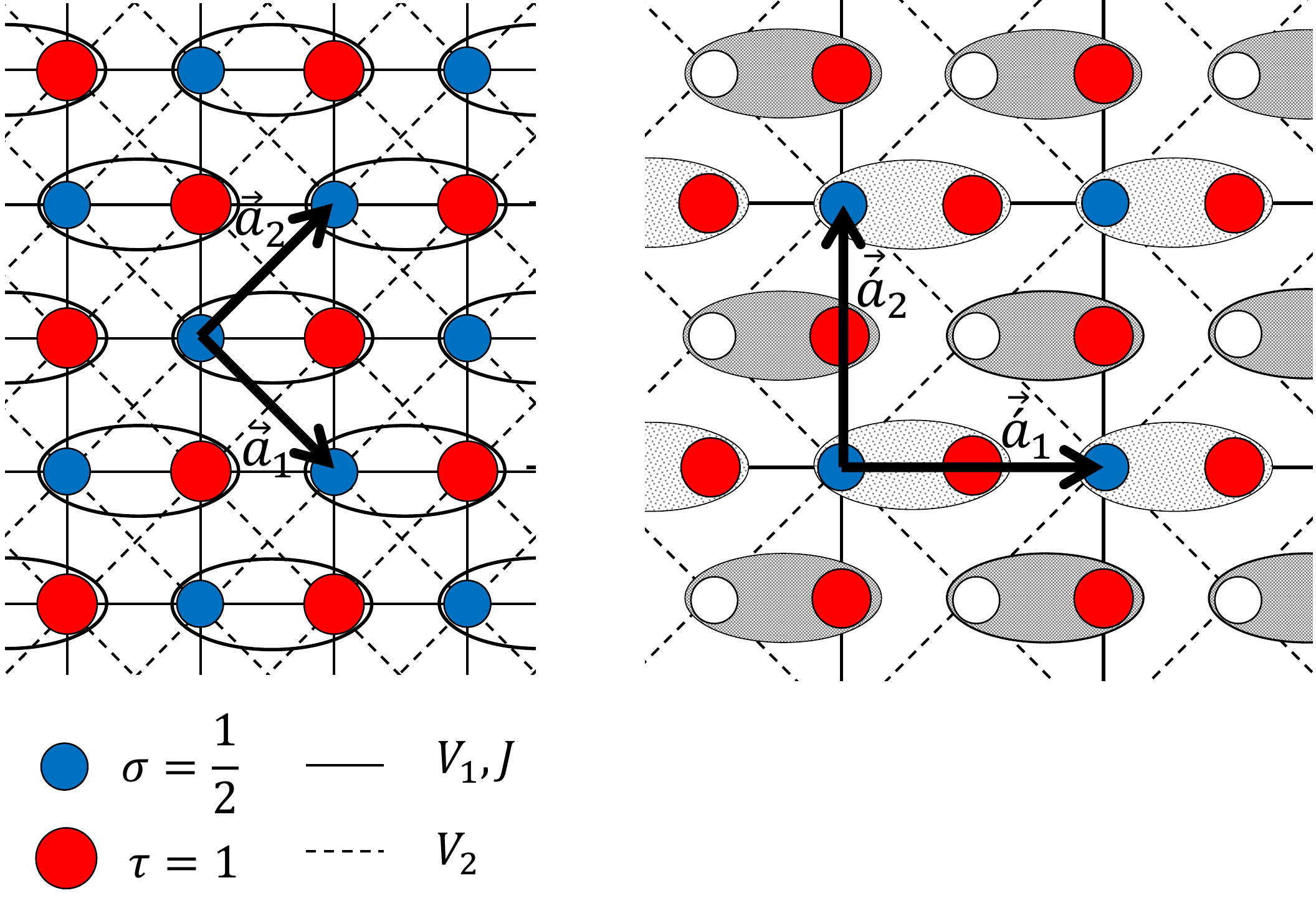}}
 \caption{(Color online)
The schematic illustration of a 2D ferrimagnetic spin-($\s,\t$) system on square lattice. Left: before the translational symmetry breaking of the Hamiltonian (\ref{Hamiltonian}), Right:
after the translational symmetry breaking in subsystem with spin $\s$. Each unit cell contains two spins $\s$ and $\t$. Before symmetry breaking the primitive vectors are $\vec{a}_1$ and $\vec{a}_2$. In the symmetry breaking phase the primitive vectors are $\vec{a^{\prime}}_1=\vec{a}_1+\vec{a}_2$ and $\vec{a^{\prime}}_2=\vec{a}_2-\vec{a}_1$. The right panel shows the checkerboard solid phase in which the translational symmetry of the subsystem with spin $\s$ is broken.} \label{fig:ferri}
\end{figure}
\begin{figure}[t]
\centerline{\includegraphics[width=80mm]{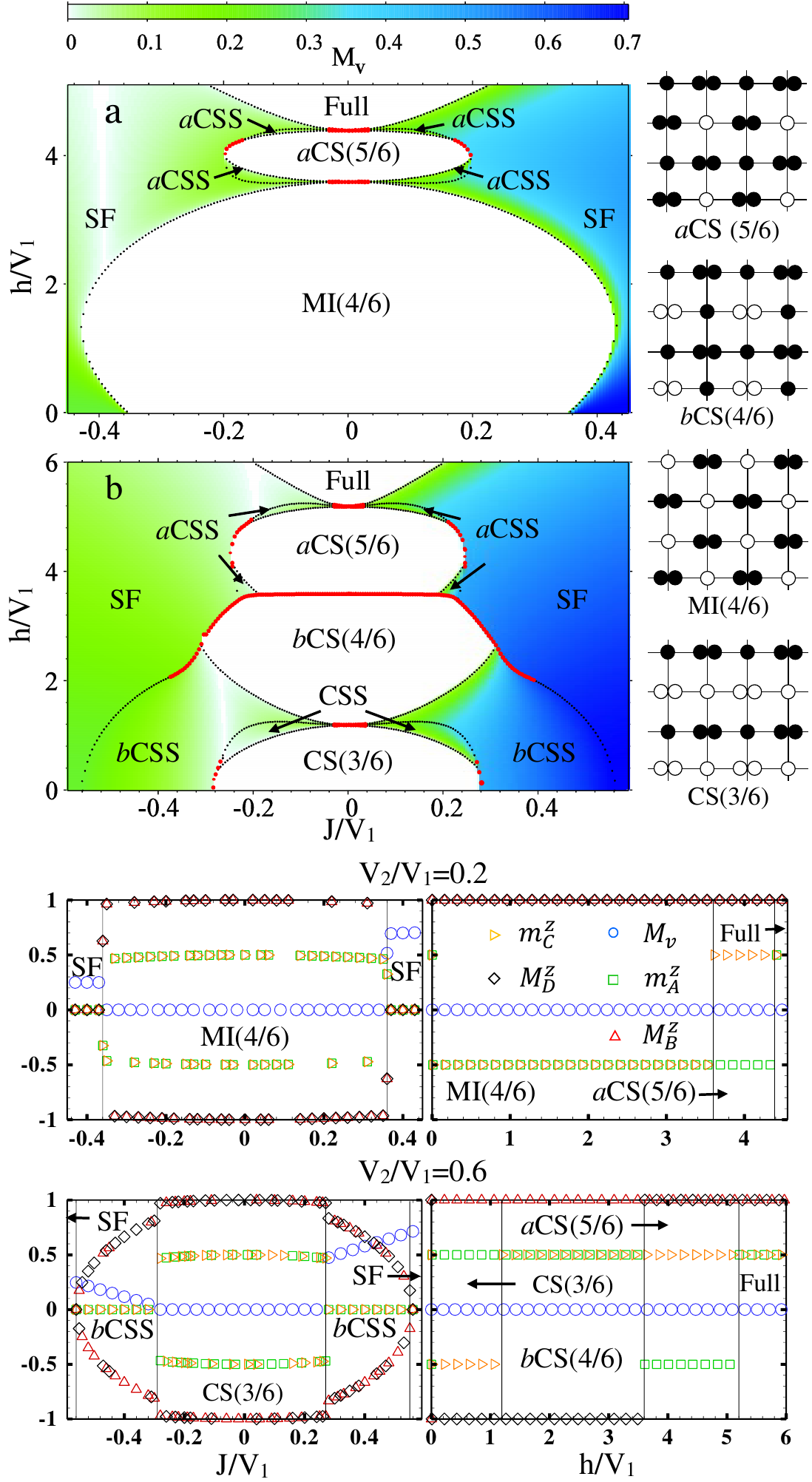}} \caption{(Color online) Top (a and
b): Ground state phase diagram of the IBM in the absence of $U$, and for the two different strengths of frustration $\frac{V_2}{V_1}$, 0.2 (a) and 0.6 (b).
Order parameters are computed using CMF-$2\times 2$
theory. The density of colors shows
amount of off diagonal order parameter: $M_{v}=((M^x_T)^2 +
(M^y_T)^2)^{1/2}$ with $M_T^{x(y)}$, the total
magnetization in $x(y)$ direction. The red(black) dotted lines show
first(second) order phase transitions.
Right: Schematic illustrations of solids and Mott insulator. Different orders are defined as in table
\ref{tab:Corders}. Bottom: The sublattices longitudinal
magnetization and the total transverse magnetization versus $J/V_1$
at $h=0$, and versus $h/V_1$ at $J=0$.} \label{phase-diag-U0}
\end{figure}

Using the Matsubara-Matsuda transformations\cite{Matsubara1956} for $a$ hardcore bosons:
\begin{equation}
\label{MM}\s_i^z=n_i^a-\frac 12,~~~\s_i^+=\dg a_i,~~~\s_i^-=a_i,
\end{equation}
and also the generalized transformations\cite{Batista2004}
for $b$ bosons:
\begin{equation}
\label{BO}\tau_j^z=n^b_j-1,~~~
\tau_j^+=\sqrt{2}\dg b_j,~~~
\tau_j^-=\sqrt{2} b_j,
\end{equation}
the Hamiltonian $H_B$ transforms to the following spin Hamiltonian:
\begin{eqnarray}
\nonumber H&=&-2J\sum_{\la i,j\ra}(\s_i^x\t_j^x+\s_i^y\t_j^y)
+U\sum_i(\tau_i^z)^2+V_1\sum_{\la i,j\ra}\s_i^z\t_j^z\\
\label{Hamiltonian}&&+V_2\sum_{\la\la
i,j\ra\ra}(\s_i^z\s_j^z+\t_i^z\t_j^z)
-\sum_i(h^{\s}\s_i^z+h^{\t}\t_i^z),
\end{eqnarray}
with the parameters $J=\sqrt 2t$, $h^{\s}=\mu^a-4V_1-4V_2$ and $h^{\t}=\mu^b-2U-2V_1-8V_2$. This Hamiltonian is nothing
but the frustrated anisotropic mixed spin-($1, 1/2$) XXZ model on a bipartite square lattice, with on-site anisotropy, in the presence of longitudinal magnetic fields $h^\s$ and $h^\t$.
Since the hardcore boson-spin transformations (\ref{MM}) and
(\ref{BO}) are isomorphic, the symmetries and physical properties
of the IBM (\ref{BH1}) and the mixed spin
Heisenberg model (\ref{Hamiltonian}) are identically the same.
Throughout this paper we consider $h^\s=h^\t=h$, which results the relation $\mu^b-\mu^a=4V_2-2V_1-2U$, between the chemical potentials of the two species.
A schematic illustration of the ferrimagnetic model (\ref{Hamiltonian}) is depicted in Fig. \ref{fig:ferri}. The small(large) filled circles are the spins $\s(\t)$. In the presence of the translational symmetry of the Hamiltonian (\ref{Hamiltonian}) the primitive vectors are $\vec{a}_1$ and $\vec{a}_2$. When the translational symmetry breaks (at least in one of the subsystems) a phase transition occurs to a checkerboard solid phase in which the lattice structure is given by the primitive translational vectors $\vec{a'}_1$ and $\vec{a'}_2$ with four basis. As an example we have illustrated in the right panel of Fig. \ref{fig:ferri} a checkerboard pattern where the translational symmetry of the subsystem with spin $\s$ is broken.

In anisotropic spin-1/2 models on square lattice with NN and NNN interactions, due to the strength of frustration, quantum fluctuations in spin direction are large enough to destroy the CSS order. In contrast, we will demonstrate that the anisotropic ferrimagnetic spin-(1, 1/2) model in Eq. (\ref{Hamiltonian}) possesses an stable CSS phase. This is in part due to the fact that each spin-1/2 is surrounded by four spins 1 which causes decreasing of quantum fluctuations. Besides the CSS phase, different solid orders and Mott insulating phase emerge in the phase diagram of the system which are not seen in the homogeneous spin $\frac{1}{2}$ models.
In following sections utilizing CMF approach we study the phase diagrams of the model (\ref{Hamiltonian}) on a square lattice.

\section{Cluster mean field theory}\label{sec:CMF}

CMF theory is an extension of the standard mean field (MF) theory in
which "{\it clusters}" of multiple sites are used as an approximate
system instead of single sites. Treating exactly the interactions
within the cluster and including the interaction of spins outside
the cluster as an effective field, one can partially take into
account fluctuations around classical ground state as well as the
effects of correlations of particles. We have generalized the CMF
approach of Yamamoto, et al \cite{PhysRevB.86.054516,
PhysRevA.85.021601, yamamoto2009correlated} which is an extension of Oguchi's method\cite{Oguchi1955} to multiple-sublattice problems, to the inhomogeneous mixed-spin model in Eq. (\ref{Hamiltonian}). We assume a background with four-sublattice structure (A and C for spins $\s$, and B and D for spins $\t$) and embed a cluster of $N_C$ sites into this background. The four-sublattice structure is expected to be emerged due to the NN and NNN interactions. Now, instead of treating the many-body problem in the whole system, we consider the effective cluster Hamiltonian:
\begin{equation}
H_C^{eff}=H_C+\sum_{i\in C}(\vec{h}_i^{eff}\cdot\vec{\s}_i+\vec{g}_i^{eff}\cdot\vec{\t}_i)
\label{C-Hamiltonian},
\end{equation}
where the interaction within cluster is given by $H_C$, the Hamiltonian in Eq. (\ref{Hamiltonian}) with $i,j \in C$,
while the interactions of spins inside the cluster with the rest of the system are included via the effective fields:
\begin{eqnarray}
\nonumber\vec{h}_i^{eff}&=&\sum_{\la i,j\ra , j\in \bar{C}}[-2J(M_j^x\hat{x}+M_j^y\hat{y})+V_1M_j^z\hat{z}]\\
\nonumber &+& V_2\sum_{\la \la i,j\ra \ra , j\in \bar{C}} m_j^z\hat{z},\\
\nonumber\vec{g}_i^{eff}&=&\sum_{\la i,j\ra , j\in \bar{C}}[-2J(m_j^x\hat{x}+m_j^y\hat{y})+V_1m_j^z\hat{z}]\\
&+& V_2\sum_{\la \la i,j\ra \ra , j\in \bar{C}} M_j^z\hat{z},
\end{eqnarray}
with $\bar{C}$ part of the system outside the cluster. The magnetizations $\vec{m}_j(=\la \vec{\s}_j\ra_{CMF})$ and
$\vec{M}_j(=\la \vec{\t}_j\ra_{CMF})$ are the expectation values within the CMF method which act as mean fields on the spins $\s$ and $\t$.
The order parameters $m_j^{x,y,z}$ and $M_j^{x,y,z}$ are calculated self-consistently as the expectation values of the spins inside the cluster.
This method reduces to the conventional MF theory for $N_C^\t=N_C^\s=1$ and becomes exact in the limit of $N_C \rightarrow \infty$.

\section{Ground state phase diagram}\label{sec:GPD}

\begin{figure}[t]
\centering
\includegraphics[width=70mm]{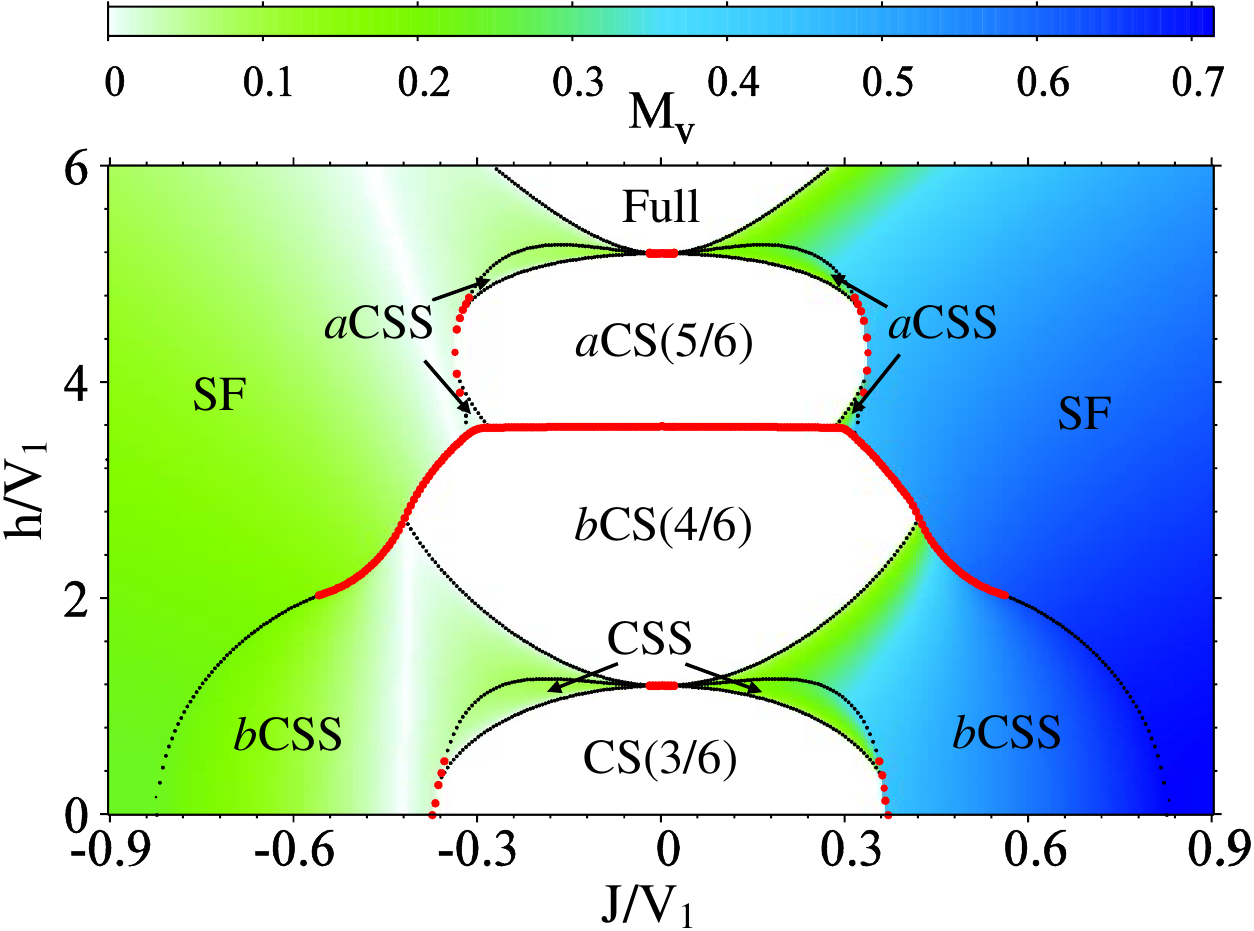}
\caption{(Color online) Ground state phase diagram of the iBH mixture for
$V_2/V_1=0.6$ and $U/V_1=-1.4$.} \label{phase-diag-U}
\end{figure}

According to the relations between the sublattices magnetizations various kinds of solid and supersolid orders are observed in the ground state phase diagram of the IBM (See Fig. \ref{phase-diag-U0} for $U=0$ and Fig. \ref{phase-diag-U} for $U=-1.4 V_1$).

In the absence of magnetic field, at $h=0$, for large values of
hopping energy $|J|$ and any strength of frustration, the IBM
is in a superfluid (SF) phase where the U(1) symmetry of
both subsystems is broken and each boson is spread out over the
entire lattice, with long range phase coherence. By decreasing $|J|$
the IBM however, behaves differently for strong and weak
frustrations (see Fig. \ref{phase-diag-U0}, bottom panels, the
behavior of the sublattices longitudinal magnetizations versus
$J/V_1$ at $h=0$). For $V_2/V_1<0.4$, at the first order transition
line: $V_2\approx-2.02 J+0.77 V_1$ (not shown) the off diagonal long
range order are suddenly destroyed and a quantum phase transition
occurs from SF to the MI(4/6) Mott insulating phase where both the
U(1) and the translational symmetries are preserved. In this phase
the average number of bosons in each unit cell is $4/6$. Increasing
$V_2/V_1$, destroys this Mott insulating phase. For $V_2/V_1\geq
0.4$, by decreasing $|J|$ the translational symmetry of the
subsystem $b$ also breaks and a phase transition from SF to the
$b$CSS supersolid phase occurs at the first order transition line:
$V_2\approx 1.53 J+0.07 V_1$, where the checkerboard solid order
emerges in the subsystem $b$ in addition to the off-diagonal one
(see table \ref{tab:Corders} for the definition of supersolid
phases). By further decreasing of $|J|$, the off diagonal order
disappears at the transition line: $V_2\approx 1.92 J+0.06 V_1$, and
the translational symmetry of subsystem $a$ also breaks and the
mixture enters the CS(3/6) solid phase. In this phase the spins 1 as
well as the spins $1/2$ are antiparallel and the average number of
bosons on each unit cell is $3/6$.
\begin{table*}
\caption{Definitions of various orders. TS is the abbreviation of the translational symmetry of the Hamiltonian.}
\label{tab:Corders}
\begin{ruledtabular}
\begin{tabular}{ccccc}
 &\multicolumn{2}{c}{Order parameters} &\multicolumn{2}{c}{broken symmetries}\\
Phases&sublattices magnetizations&total magnetization $M_v$& subsystem $a$ & subsystem $b$\\
\hline
SF& $m_A^z=m_C^z$, ~$M_B^z=M_D^z$& $\neq 0$ & U(1) & U(1)
\\
MI(4/6)& $m_A^z=m_C^z$, ~$M_B^z=M_D^z$& 0 & - & -\\
Full& $m_A^z=m_C^z=1/2$, ~$M_B^z=M_D^z=1$& 0 & - & -\\
$a$CS(5/6)& $m_A^z=-m_C^z$, ~$M_B^z=M_D^z$& 0 & TS & -\\
$b$CS(4/6)& $m_A^z=m_C^z$, ~$M_B^z=-M_D^z$& 0 & - & TS\\
CS(3/6)& $m_A^z=-m_C^z$, ~$M_B^z=-M_D^z$& 0 & TS & TS\\
$a$CSS& $m_A^z\neq m_C^z$, ~$M_B^z=M_D^z$& $\neq 0$& TS, U(1) &U(1)
\\
$b$CSS& $m_A^z=m_C^z$,~$M_B^z \neq M_D^z$& $\neq 0$& U(1) & TS, U(1)
\\
CSS& $m_A^z \neq m_C^z \neq M_B^z \neq M_D^z$& $\neq 0$ & TS, U(1)& TS, U(1)
\\
\end{tabular}
\end{ruledtabular}
\end{table*}

In the presence of magnetic field, for $h\neq 0$,
depending on the strength of frustration, various kinds of solid
order appear in the phase diagram of the IBM. We have
plotted in Fig. \ref{phase-diag-U0} the phase diagram of the mixture
for the two strengths of frustration, $V_2/V_1=0.2$ and
$0.6$. For the weak frustration $V_2/V_1=0.2$, in a symmetric region
around $J=0$, at small and moderate magnetic fields the system prefers to be in the MI(4/6) phase.
 By increasing the magnetic field, the translational symmetry of the subsystem $a$ is broken and the spins $1/2$ on one of the sublattices
A or C flip in the direction of the magnetic field and the mixture
enters the $a$CS(5/6) solid phase (For the definitions of solid
orders, see table \ref{tab:Corders} and the schematic pictures in
the right column of Fig. \ref{phase-diag-U0}). In this phase the
average number of bosons on each unit cell is $5/6$. By increasing
$V_2/V_1$, the antiferromagnetic $V_2$ interactions try to make the
spins $1$ antiparallel as well as spins $1/2$. For $V_2/V_1=0.6$
around $J=0$, the translational symmetry of both subsystems breaks
and the CS(3/6) solid order emerges in the system. By increasing the
magnetic field, the translational symmetry of the subsystem $a$ is
restored and a phase transition occurs to the $b$CS(4/6) where the
average number of bosons on each unit cell is 4/6. By further
increasing of the magnetic field, the $a$CS(5/6) solid also appears
in the phase diagram of the model just below the saturation field.
As the solid phases possess different broken symmetries, we expect
the transitions between solid phases to be first order which are
illustrated with red dotted lines in the phase diagrams.

Besides the superfluid, solids and Mott insulator, various supersolids also appear in the phase diagram of the mixture (see the definition of supersolid orders in table
\ref{tab:Corders}). For the whole range of
 $V_2/V_1$, in the two narrow regions on
the top and bottom sides of the $a$CS(5/6) solid phase, the spins tend
to lie in the plane perpendicular to the magnetic field. In these
regions, the system exhibits the $a$CSS supersolid phase in which
both diagonal (solid $a$CS(4/6)) and off diagonal long range orders coexist
in the system. Increasing the hopping parameter $|J|$, the translational symmetry of the subsystem $a$ restores and the
$a$CS(5/6) solid order disappears where a phase transition occurs from $a$CSS to the SF phase. For larger values of $V_2/V_1$, two other supersolid orders, the $b$CSS and the CSS phases, also appear
in the phase diagram at small magnetic fields around the CS(3/6)
solid phase (see Fig. \ref{phase-diag-U0}-b).
Phase transitions from the $a$CSS and $b$CSS to the SF are of first or second order, depending on the values of $h$ and $|J|$.
All these first and second order phase transitions, are attributed to the behavior of the low energy spin wave excitation which will be discussed in section \ref{sec:LSW}.
\begin{figure}[t]
\centering
\includegraphics[width=85mm]{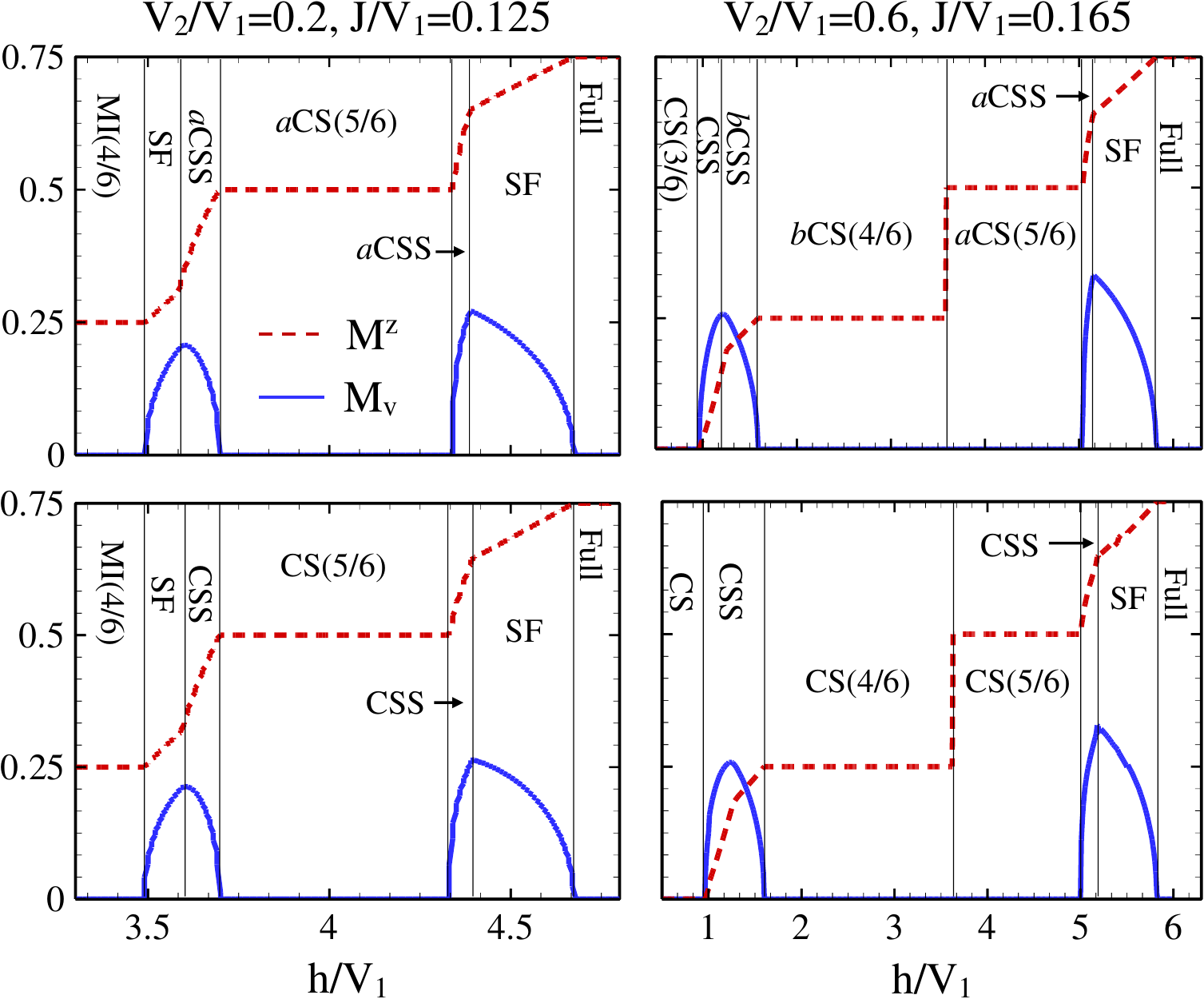}
\caption{Diagonal and off diagonal order parameters, computed using
CMF theory with $2\times 2$ (top) and $2\times 4$ (bottom) clusters,
for the two different strengths of frustration: $V_2/V_1=0.2$ and
$0.6$, and the two values of hopping parameter: $J/V_1=0.125$ and
$0.165$ where all phases appear in the phase diagram by increasing $h$. According to the CMF-$2\times 4$ results, quantum fluctuations
convert the $a$CSS and $b$CSS phases into the CSS phase. The CSS, SF and MI(4/6) phases are not changed by quantum fluctuations.}
\label{fig:order-parameter}
\end{figure}
\begin{figure}[h]
\centering
\includegraphics[width=60mm]{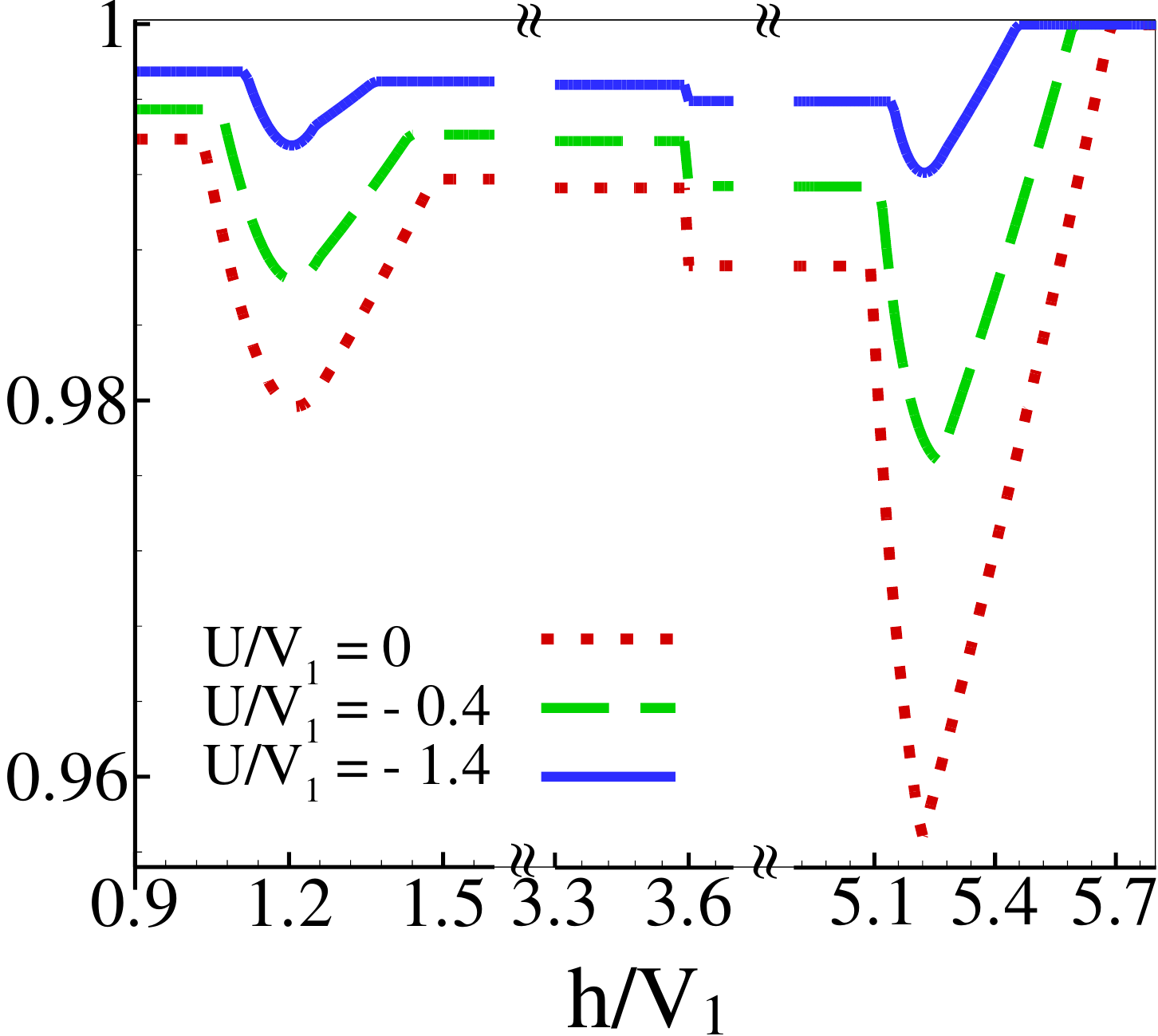}
\caption{The order parameter $Q^z$ versus
magnetic field for different values of on-site interaction, at
$V_2/V_1=0.6$ and $J/V_1=0.14$.} \label{fig:Q}
\end{figure}
\begin{figure*}[t]
\centering
\includegraphics[width=155mm]{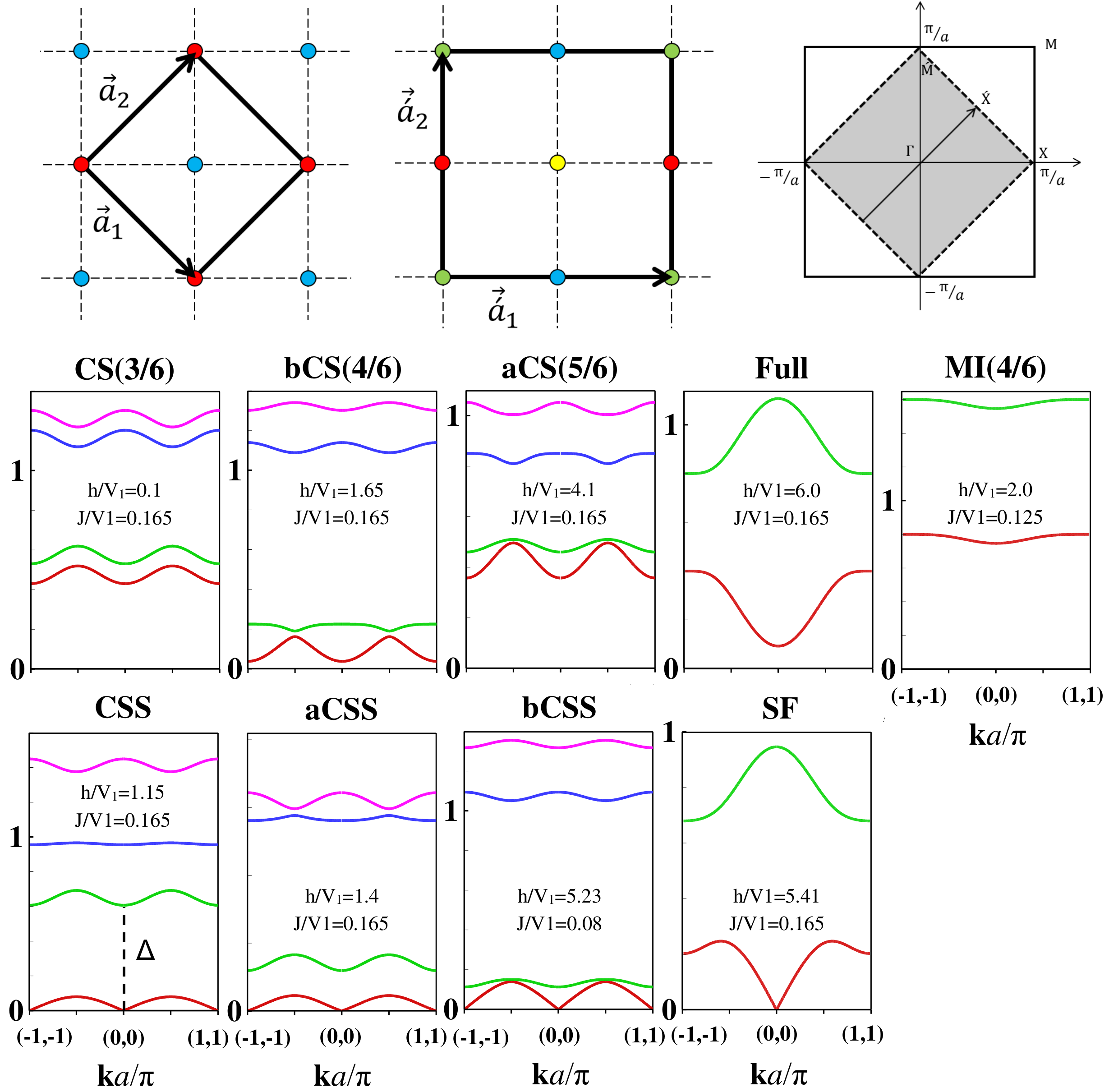}
\caption{(Color online) Excitation spectra in various phases of the
IBM.  Number of excitation modes reflects the number of
sublattices in each phase. Top-left: 2D lattice with primitive
vectors $\vec{a}_1=a \hat{x}$ and $\vec{a}_2=a\hat{y}$ for the
MI(4/6), SF and Full phases. Top-center: 2D lattice with primitive
vectors $\vec{a}_1=a(\hat{x}+\hat{y})$ and
$\vec{a}_2=a(-\hat{x}+\hat{y})$ for the solid and supersolid phases
where the original lattice symmetry is broken. Top-right: the
unfolded and folded Brillouin zones. Middle and bottom: spin wave
excitations in all phases in $k_x=k_y$ direction of the unfolded
Brillouin zone. The roton gap ($\Delta$) varies in each supersolid
phase. All plots are for $V_2/V_1=0.6$ except MI(4/6) which is for
$V_2/V_1=0.2$. } \label{fig:SW-Excitations}
\end{figure*}

In order to see the effects of quantum fluctuations we investigate
the behavior of both the diagonal and off diagonal order parameters
considering clusters with larger sizes in CMF theory. Employing
clusters of 8 spins (CMF-$2\times 4$), we have computed the
sublattices longitudinal and transverse magnetizations for different
values of $h$ and $J$. We found out that the
quantum fluctuations convert the $a$CSS and $b$CSS phases to the CSS phase. Actually, competition between
NN and NNN interactions causes the $a$CS(5/6) and $b$CS(4/6) solids
transform respectively to the CS(5/6) and CS(4/6) solids in which
there is no relation between the sublattices longitudinal
magnetizations, but the occupation number of each unit cell is
conserved. These effects are clearly seen in Fig.
\ref{fig:order-parameter}-bottom, in the behavior of the
total magnetizations for $V_2/V_1=0.2$ at line $J/V_1=0.125$, and
for $V_2/V_1=0.6$ at line $J/V_1=0.165$, where all kinds of orders
appear in the system by increasing $h$. The MI(4/6) insulator and the
CSS supersolid are however stable and quantum fluctuations
cannot destroy these orders. This is in contrast with the standard
$V_1-V_2$ hardcore Bose Hubbard model\cite{PhysRevB.86.054516, hebert2001quantum, PhysRevLett.84.1599, PhysRevB.51.8467, dang2008vacancy, schmid2002finite} on square lattices in which the
CSS phase is unstable against quantum fluctuations, and the presence of long range dipole-dipole interactions between hardcore bosons or long range hopping terms are necessary for the stability of the CSS phase\cite{PhysRevB.86.054516, PhysRevLett.104.125301, doi:10.1143/JPSJ.80.113001}.
Actually, due to the intrinsic difference in the nilpotency condition between $a$ and $b$ bosons ($(a_i^\dagger)^2=0$ and $(b_j^\dagger)^3=0$, where $i$ and $j$ are nearest neighbor sites) quantum
fluctuations are not strong enough to destroy the MI(4/6) and CSS
phases. The stability of the CSS phase can be attributed to the large
roton-like energy gap in the low energy spin wave excitation
spectrum which will be discussed in Sec. \ref{sec:LSW}. In this section we will
obtain the amount of quantum fluctuations in terms of the
number of spin waves and show that the results of CMF-$2\times 4$
are verified by LSW theory.

As the $b$ bosons satisfy the condition $(b_j^\dag)^3=0$, one can put up two $b$ particles on each lattice site. This motivates us to investigate the behavior of the pair superfluid order parameter $\la (b_j^\dag)^2\ra$ in all phases. In the spin language this parameter is equivalent to $\frac 12\la(\t_j^+)^2\ra$.
According to our CMF results (not shown) we find that the pair superfluidity order parameter is zero in the whole range of parameter space. This means that although $(b_j^\dag)^3=0$, but no pairing occurs in the system. We have also computed the order parameter $Q^z=\frac 12\la(\t_B^z)^2+(\t_D^z)^2\ra$ in different phases. Actually, investigation of the behavior of $Q^z$ gives more intuitions on the properties of solid and Mott insulating phases appeared in the phase diagram. This order parameter is almost 1 in all solid and Mott insulating phases (see Fig. \ref{fig:Q}). This implies that the local Hilbert space basis for $b$ particles is given by the states: $\ket 0$ and $\ket 2$, and consequently the effective Hilbert space dimensions for $b$ particles is two. This fact has been already shown in the schamtic pictures of solid and Mott insulating phases in the right panel of Fig. \ref{phase-diag-U0}.

In the presence of the attractive on-site interaction, at $U\neq 0$, although
the superfluidity order parameter decreases with increasing the on-site interaction,
there is no considerable changes in the nature of the phases in comparison with the cases of $U=0$ (see Fig. \ref{fig:Q} and \ref{phase-diag-U}).
The attractive on-site interaction between $b$ particles causes these particles prefer to be at the same site to minimize the interaction energy.
This leads to the stability of all the phases at larger values of $|J|$, and therefore to the shift of the phases' boarders to the larger values of $|J|$ in the presence of $U$.
\begin{figure*}
\centering
\includegraphics[width=160mm]{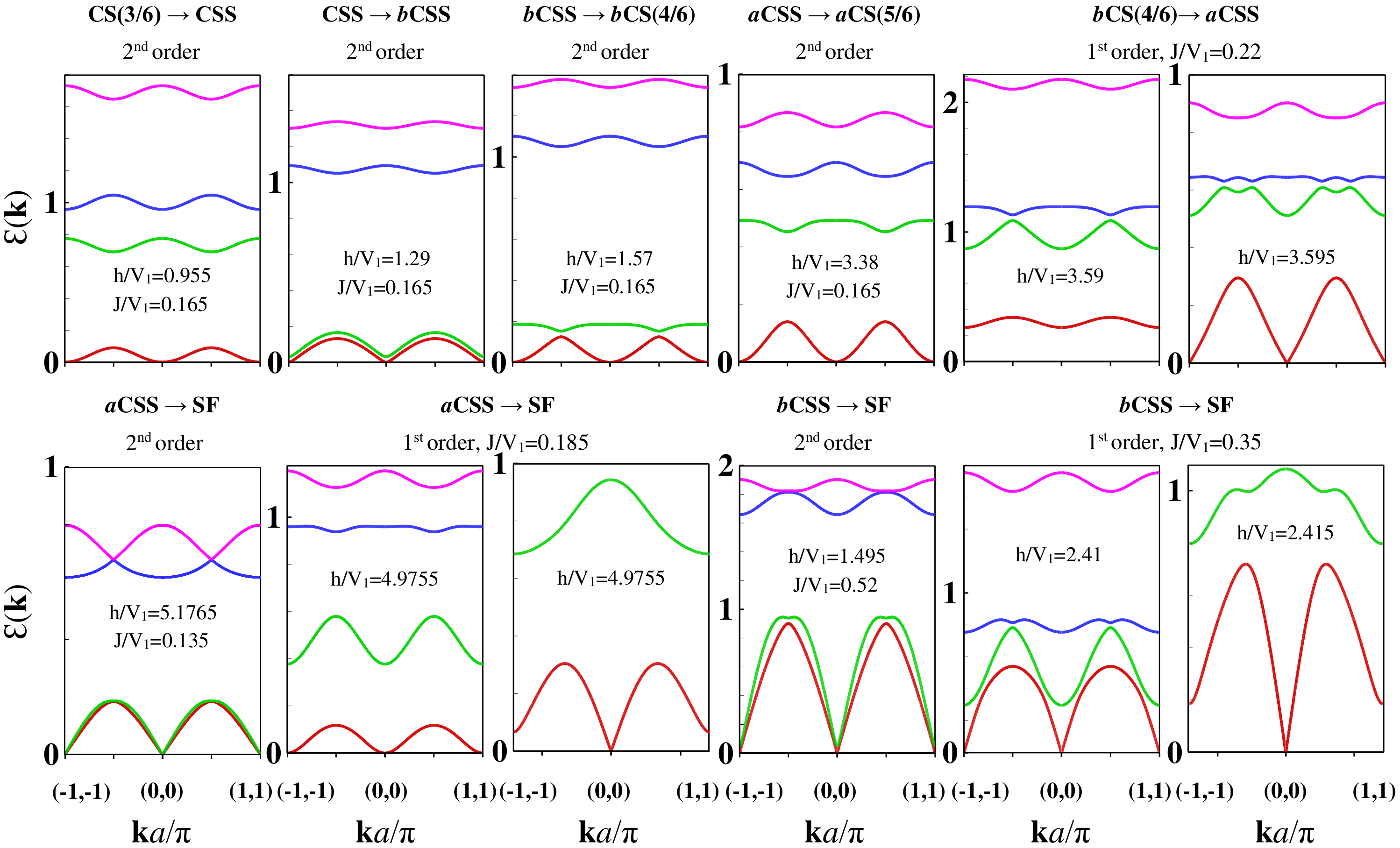}
\caption{(Color online) Excitation spectra at various first and
second order phase transitions
for $V_2/V_1=0.6$. In the second order supersolid-solid phase transition (phase transition from $a$CSS to $a$CS(5/6), from $b$CSS to $b$CS(4/6) and from CSS to CS(3/6)) the linear dispersions around $\kv=(0,0)$ and $\kv=(\pi/a,\pi/a)$ softens to quadratic ones and become gapped in solid phases.
In second order phase transitions from $a$CSS to SF, and from $b$CSS to SF number of excitations changes and the amount of the roton gap varies continuously. At first order
$b$CSS-SF phase transition the roton gap decreases suddenly. At first
order $b$CS(4/6)-$a$CSS phase transition the lowest gapped mode abruptly
touches zero and the quadratic dispersion changes to a linear one
around $\kv=(0,0)$ and $\kv=(\pi/a,\pi/a)$.}
\label{fig:excit-transition}
\end{figure*}
\begin{figure*}
\includegraphics[width=150mm]{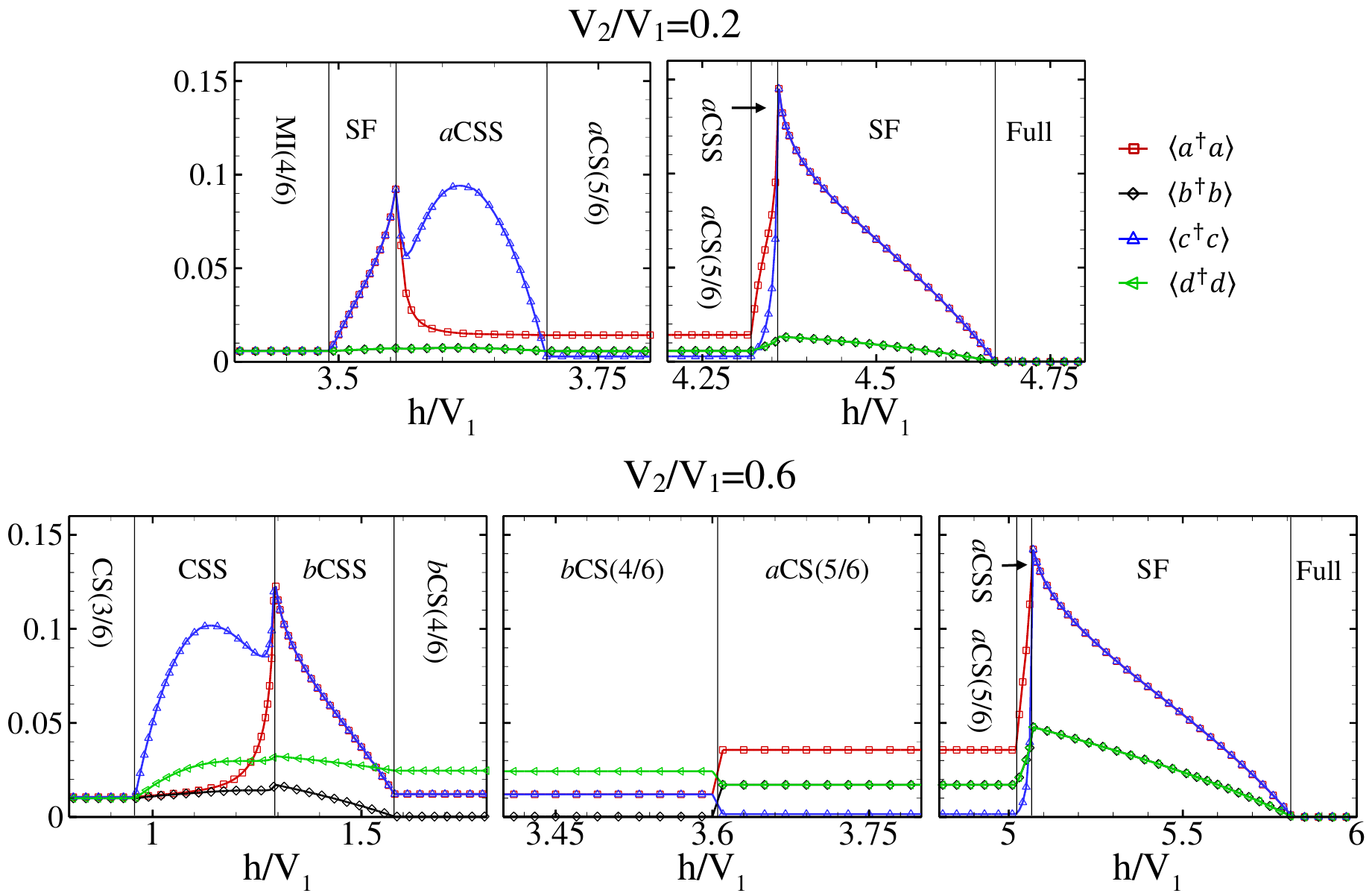}
\caption{(Color online) Number of HP bosons around mean field ground state. Top: $V_2/V_1=0.2$, at line $J/V_1=0.126$. Bottom: $V_2/V_1=0.6$, at line $J/V_1=0.166$.} \label{fig:SW-fluc}
\end{figure*}

\section{Linear spin wave theory}\label{sec:LSW}

In this section, utilizing LSW theory we obtain the excitation
spectra of the mixed spin model in Eq. (\ref{Hamiltonian}). From these spectra besides the strength
of quantum fluctuations around the MF ground states, one can find
the boundaries of the stability of the mean field phases.
Furthermore, by investigating the behavior of the low energy spin
wave dispersions at phase
transitions one can figure out the reason of all first and second
order phase transitions. Before starting the spin wave approach we
implement a unitary transformation on the spin Hamiltonian in Eq.
\ref{Hamiltonian} and perform the following rotations on all $\s$
and $\t$ spins;
\begin{eqnarray}
\nonumber \begin{centering}
\begin{pmatrix}
\tilde{\s}^x_i \\
\tilde{\s}^y_i \\
\tilde{\s}^z_i
\end{pmatrix}
=
\begin{pmatrix}
\cos\theta_i\cos\phi_i & -\cos\theta_i\sin\phi_i & -\sin\theta_i \\
-\sin\phi_i & \cos\phi_i & 0 \\
\sin\theta_i\cos\phi_i & \sin\theta_i\sin\phi_i & \cos\theta_i
\end{pmatrix}
\begin{pmatrix}
\s^x_i \\
\s^y_i \\
\s^z_i
\end{pmatrix}
\end{centering},
\end{eqnarray}
\begin{eqnarray}
\nonumber\begin{centering}
\begin{pmatrix}
\tilde{\t}^x_j \\
\tilde{\t}^y_j \\
\tilde{\t}^z_j
\end{pmatrix}
=
\begin{pmatrix}
\cos\vartheta_j\cos\varphi_j & -\cos\vartheta_j\sin\varphi_j & -\sin\vartheta_j \\
-\sin\varphi_j & \cos\varphi_j & 0 \\
\sin\vartheta_j\cos\varphi_j & \sin\vartheta_j\sin\varphi_j & \cos\vartheta_j
\end{pmatrix}
\begin{pmatrix}
\t^x_j \\
\t^y_j \\
\t^z_j
\end{pmatrix}
\end{centering},
\end{eqnarray}
where $\cos\theta_i=\langle\s_i^z\rangle/\s$,
$\tan\phi_i=\langle\s_i^y\rangle/\langle\s_i^x\rangle$,
$\cos\vartheta_j=\langle\t_j^z\rangle/\t$ and
$\tan\varphi_j=\langle\t_j^y\rangle/\langle\t_j^x\rangle$. Here,
$\langle\dots\rangle$ denotes the expectation value on the MF ground
state of the Hamiltonian in Eq. (\ref{Hamiltonian}). The rotated
spin Hamiltonian is expressed in terms of the new bosonic operators
$\hat{a}$, $\hat{b}$ with the following Holstein-Primakoff (HP)
transformations:
\begin{eqnarray}
\nonumber\tilde{\s}_i^z& =& \s -  \hat{a}_i^\dag \hat{a}_i,\\
\nonumber\tilde{\s}_i^+&=&\sqrt{2\s-\hat{a}_i^\dag \hat{a}_i}~{\hat{a}_i}\approx\sqrt{2\s}~\hat{a}_i,\\
\tilde{\s}_i^-&=&\hat{a}_i^\dag\sqrt{2\s-\hat{a}_i^\dag
\hat{a}_i}\approx\sqrt{2\s}~\hat{a}_i^\dag,
\label{HP1-transformation}
\end{eqnarray}
and
\begin{eqnarray}
\nonumber \tilde{\t}_j^z& =&\t- \hat{b}_j^\dag \hat{b}_j,\\
\nonumber\tilde{\t}_j^+&=&\sqrt{2\t-\hat{b}_j^\dag \hat{b}_j}~{\hat{b}_j}\approx\sqrt{2\t}~\hat{b}_j,\\
\tilde{\t}_j^-&=&\hat{b}_j^\dag\sqrt{2\t-\hat{b}_j^\dag
\hat{b}_j}\approx\sqrt{2\t}~\hat{b}_j^\dag.
\label{HP2-transformation}
\end{eqnarray}
The spin wave Hamiltonian has the following form;
\begin{equation}
\label{SWH}\tilde{H}=E_0 + H^\prime,
\end{equation}
where $E_0$ is the classical MF energy and $H^\prime$
consists of bilinear terms in HP boson operators. This part yields, after
diagonalization, the excitation spectra in each phase (For the details of diagonalization see the
appendix \ref{appendix}).

From general symmetry analysis, the off diagonal order parameter
manifold has U(1) freedom to rotate the transverse spin order around
the magnetic field direction. In the superfluid phase, the U(1)
symmetry is spontaneously broken and a gapless Goldstone mode with a
roton-like minimum appears in the excitation spectra. The slope of
the line connecting the origin of the $\ve-\kv$ plane with this
minimum is proportional to the critical velocity of the superfluid
and the energy of this minimum is the roton energy gap. Upon
approaching the transition (second order) from the superfluid side
the roton energy and consequently the critical velocity decrease to
zero. At the same time the superfluid order parameter remains finite
through the supersolid transition. Inside supersolid phases due to
the translational symmetry breaking the spatial periodicity is
doubled, and the Brillouin zone becomes smaller. So half of the
excitation spectrum is folded back to the point $\kv=(0,0)$. This
second branch acquires a gap, with a quadratic minimum above it, and
so the critical velocity will continuously climb back to finite
values. The critical velocity in each supersolid phase varies by
changing the magnetic field. By comparison of the maximum critical
velocities in each phase we have found out that they satisfy the
inequality $v_{aCSS}<v_{bCSS}<v_{CSS}$.

We have plotted in Fig. \ref{fig:SW-Excitations} the spin wave
excitation spectra of all phases. Number of excitation modes and
their behavior depend on the number of sublattices as well as their
longitudinal and transverse magnetizations. According to the
Brillouin zone folding, the $k_x=k_y$ direction in the unfolded zone
corresponds to the $k_x$ direction in the folded one (See Fig.
\ref{fig:SW-Excitations}), and so the points $\kv=(0,0)$ and
$(\pi/a,\pi/a)$ are equivalent. In the SF phase and all the
supersolid phases the lower excitation has linear dispersion around
the points $\kv=(0,0)$. Investigation of the amount of the roton gap
in supersolid phases helps us to figure out their stability in the
presence of quantum fluctuations. The small roton gap in the $a$CSS
and $b$CSS phases causes fluctuations annihilate low energy rotons
and convert the $a$CSS and $b$CSS phases to the CSS one.

In a solid phase there is no Goldstone zero mode and all excitations
are gapped. The lowest gapped excitation spectrum has quadratic
dispersion ($k^2$) around $\kv=(0,0)$. In CS(3/6) solid and MI(4/6)
Mott insulating phases we have found a relation between the
excitation energies separation and the magnetic field. In MI(4/6)
phase the two energy bands are related by:
\begin{eqnarray}
\nonumber &&|\ve^2(\kv)-\ve^1(\kv)|=-\frac{h}{V_1}+1.2,~~~0<\frac{h}{V_1}<1.2,\\
&&|\ve^2(\kv)-\ve^1(\kv)|=\frac{h}{V_1}-1.2,~~~~~1.2<\frac{h}{V_1},
\end{eqnarray}
and in the CS(3/6) phase, the four energy bands have the
relation
$|\ve^1(\kv)-\ve^2(\kv)|=|\ve^3(\kv)-\ve^4(\kv)|=\frac{h}{V_1}$,
where $\ve^1$ and $\ve^2$, and $\ve^3$ and $\ve^4$ are the two
branches with the same energy behaviors. By increasing magnetic
field the two lowest energy bands repel each other causing the
energy of the lowest mode decreases and touches zero at $\kv=(0,0)$,
when a second order phase transition
to the CSS phase occurs in the system.

In order to find out the reason of all first and second order phase transitions in the ground state phase diagram of our IBM, we have also plotted in Fig. \ref{fig:excit-transition} the excitation spectra at the phase transition between different phases.
The abrupt and smooth changes in the behavior of the low energy excitation modes and the roton gap are the reason of first and second order phase transitions, respectively.
In the second order supersolid-solid phase transitions the roton minimum at $\kv=(0,0)$ disappears and the low energy mode softens around this point and becomes gapped in solid phases.

Quantum fluctuations around MF ground state are given by
\begin{eqnarray}
\begin{split}
\la a^\dag a\ra&=\la \s^z_A\ra_{MF}-\la \s^z_A\ra_{SW},\\
\la c^\dag c\ra&=\la \s^z_C\ra_{MF}-\la \s^z_C\ra_{SW},\\
\la b^\dag b\ra&=\la \t^z_B\ra_{MF}-\la \t^z_B\ra_{SW},\\
\la d^\dag d\ra&=\la \t^z_D\ra_{MF}-\la \t^z_D\ra_{SW},
\label{Eq:Fluctuations2}
\end{split}
\end{eqnarray}
which are the number of $\hat{a}$, $\hat{b}$, $\hat{c}$, and
$\hat{d}$ HP bosons. Here, $\la \s^z_{A,C}\ra_{MF}$ and $\la
\t^z_{B,D}\ra_{MF}$, and $\la \s^z_{A,C}\ra_{SW}$ and $\la \t^z_{B,D}\ra_{SW}$ are the MF and LSW sublattices' magnetizations, respectively.
We have plotted in Fig. \ref{fig:SW-fluc} the number of HP
bosons versus magnetic field, for the two different strengths of
frustration $V_2/V_1=0.2$ and $0.6$. Larger frustration in the case
of $V_2/V_1=0.6$ due to the competition between NN and NNN
interactions causes stronger quantum fluctuations for $V_2/V_1=0.6$
in comparison with the case of $V_2/V_1=0.2$. For all strengths of
frustration the number of spin waves increases in the vicinity of
the transition points which is a result of the strong quantum
fluctuations at phase boundaries. Strong quantum
fluctuations in the $a$CSS and $b$CSS phases is the reason of the
converting of these phases to the CSS phase in the CMF-$2\times 4$
results. As we have already mentioned, this instability is also
reflected from the behavior of the roton gap in the low energy
excitation spectrum.

The small number of $\hat{b}$ and $\hat{d}$ HP bosons in comparison
with the number of $\hat{a}$ and $\hat{c}$ is an indication of the
weaker quantum fluctuations in the sublattices $B$ and $D$. The
maximum values of $\la a^\dag a\ra(=\la c^\dag c\ra)$ and $\la b^\dag
b\ra(=\la d^\dag d\ra)$ in CSS phases reach respectively about 30 and
5 percent of the classical values of the spin lengths $\s=1/2$ and
$\t=1$. This means that the prediction for the ground states within
CMF-$2\times 4$ theory are reliable. It is worth to compare the
amount of quantum fluctuations around the MF ground states of our
IBM with the standard Bose Hubbard one. The maximum value of
quantum fluctuations in CSS phase of the standard Bose Hubbard is
about 77 percent of the classical spin length\cite{PhysRevB.86.054516} which is much larger than the case of IBM, and cause the CSS phase of the standard Bose Hubbard model, predicted by MF to be unreliable.

\section{Finite temperature phase diagram}\label{sec:TPD}

At zero temperature, in the superfluidity state each boson is spread
out over the entire lattice, with long range phase coherence.  At
finite temperature, the superfluid density is suppressed and the system undergoes a transition to a thermal insulating phase with varying filling factor.
The thermal insulator (TI) is a weak Mott insulator in the sense that it preserves both the translational and the U(1) symmetries.
\begin{figure}
\centering
\includegraphics[width=70mm]{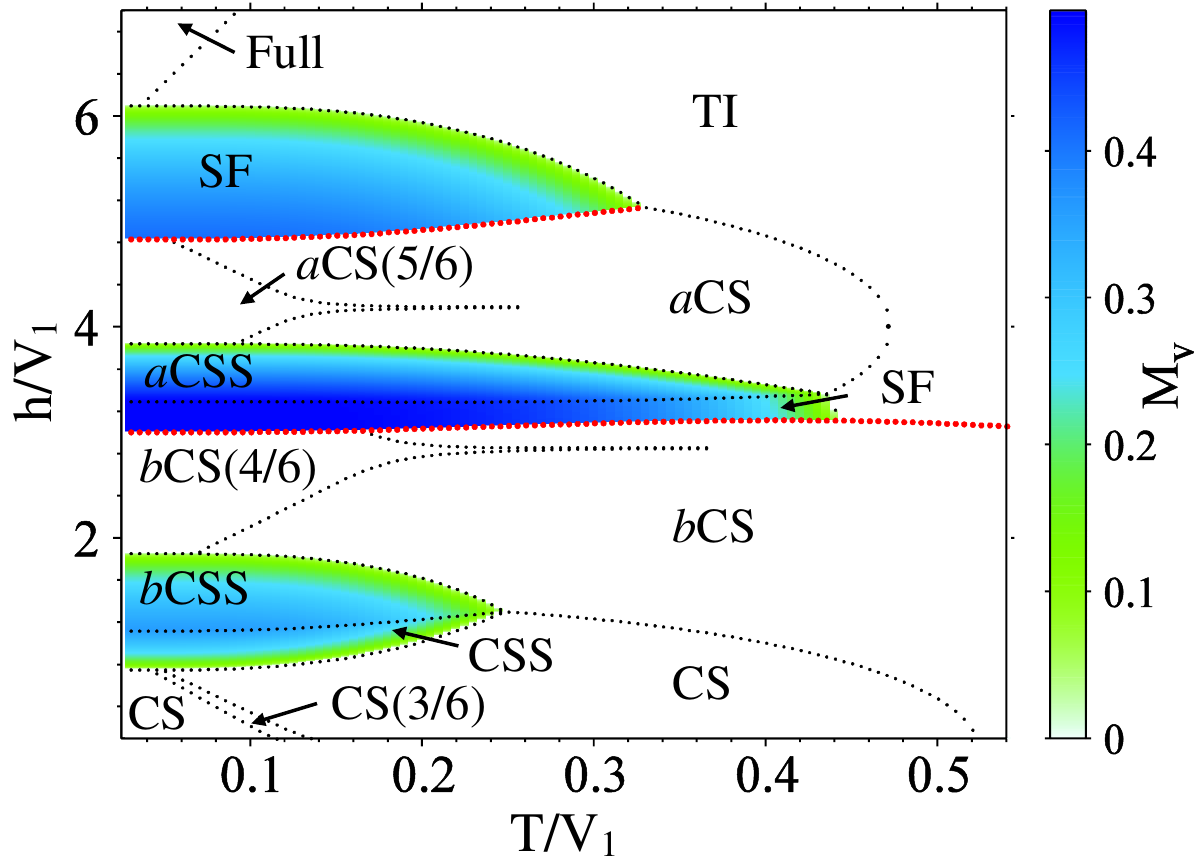}
\includegraphics[width=70mm]{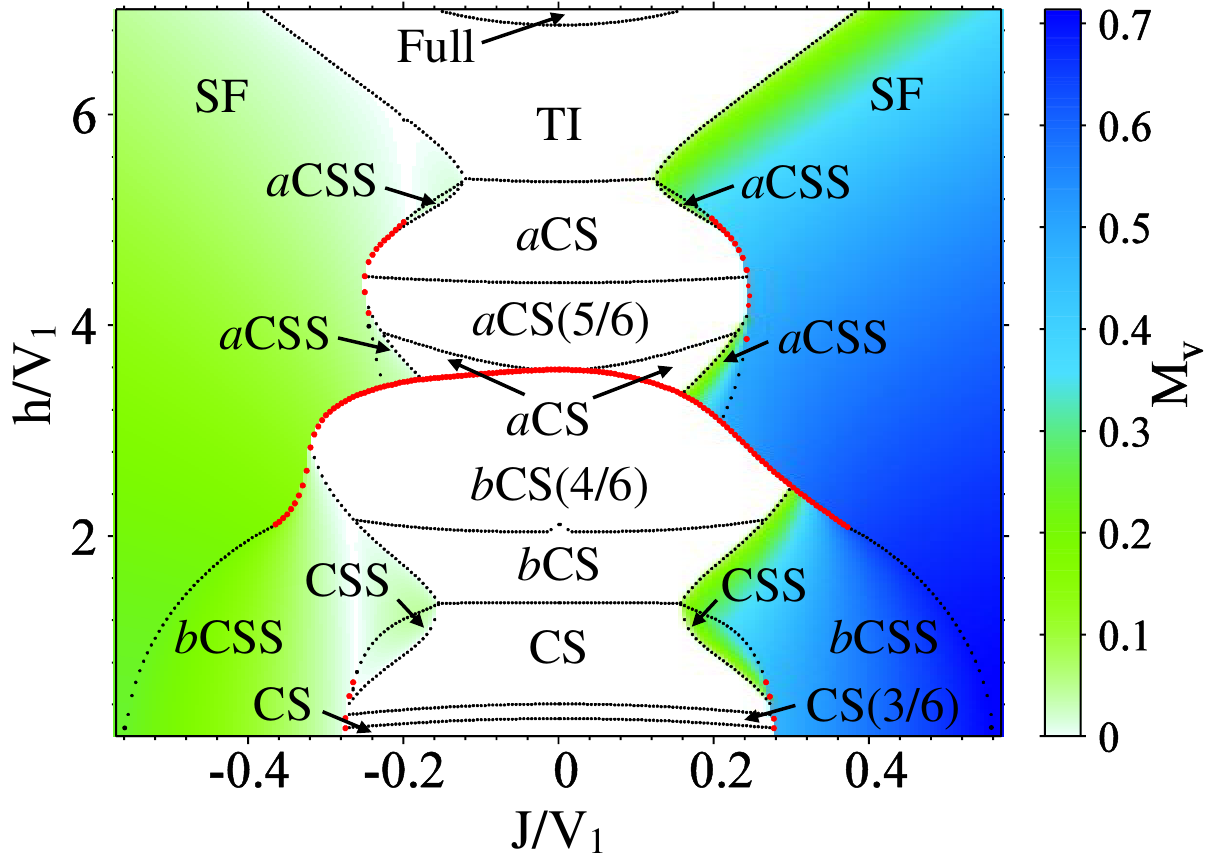}
\caption{(Color online) Top: CMF $T-h$ phase diagram of the IBM for $V_2/V_1=0.6$ at line $J/V_1=0.22$, where all phases emerge by increasing the magnetic field.
The SF-TI transition temperature for the two values of magnetic fields $h=3.22 V_1$ and $h=5.18 V_1$ are respectively $0.44 V_1$ and $0.32 V_1$.
The CSS-CS transition temperature at $h/V_1=1.30$ is $0.25 V_1$.
Bottom: CMF $J-h$ phase diagram for the frustration strength $V_2/V_1=0.6$ and $T/V_1=0.1$.}
\label{fig:Temp-phase-diag}
\end{figure}
In the presence of temperature, at $T\neq 0$, in the solid phases
the plateaus' width on longitudinal magnetization curves decreases
gradually by increasing temperature and disappears eventually at a
transition temperature which depends on the strength of frustration,
the hopping energy $J$ and the magnetic field $h$. For example, for
the  strength of $V_2/V_1=0.6$ and the hopping energy $J/V_1=0.22$,
as it is clearly observed from the $T-h$ phase diagram of the IBM (see the top of Fig. \ref{fig:Temp-phase-diag}), the two
solid phases $a$CS(5/6) and $b$CS(4/6) with constant average number
of bosons 5/6 and 4/6 survive at low temperature however, by
increasing temperature these regions become narrower and finally
disappear at the critical temperature $T_{c_1}\sim 0.26 V_1$ and
$T_{c_2}\sim 0.37 V_1$, where the mixture has a phase transition to
the $a$CS and $b$CS phases, respectively. In these phases number of
particles are not fixed in each unit cell. In the CSS phase, both
the diagonal and the off diagonal long range orders tend to be
destroyed by thermal fluctuations, however in comparison with the
superfluidity order the CS order is more robust. By increasing
temperature the superfluidity order parameter vanishes at a
transition temperature where the CSS-CS phase transition occurs.
(The value of this transition temperature for $V_2/V_1=0.6$ is
reported in the caption of Fig. \ref{fig:Temp-phase-diag}). The CSS
phase persists at finite temperatures comparable with the
interaction energies of boson.

In order to see the effects of temperature on the $J-h$ phase
diagram of our IBM, we have also plotted in Fig.
\ref{fig:Temp-phase-diag} the $J-h$ phase diagram of the mixture for
the frustration parameter $V_2/V_1=0.6$, at the finite temperature
$T/V_1=0.1$. In the presence of temperature, in addition to the ground state phases, several solid orders
like CS, $a$CS and $b$CS also appear in the phase diagram of the
mixture. Moreover in the region below the Full solid phase the mixture experiences the TI phase which both the U(1) and the translational symmetries of the original Hamiltonian are preserved.
This phase is not seen in the ground state phase diagram and is a result of thermal fluctuations.

\section{Summary and conclusion}\label{sec:summary}

We have introduced an inhomogeneous hardcore bosonic model composed of
two kinds of boson with different nilpotency conditions and have
shown that the model is an appropriate ground for searching various
supersolid orders. By generalizing the cluster mean field theory to
the IBM, we have studied both the ground state and the temperature
phase diagram of the model on a 2D square lattice. We have found
that in addition to the superfluidity phase, various kinds of solid
and supersolid emerge in the phase diagram of the inhomogeneous mixture. We
have also found that for small strengths of frustration the system
possesses a Mott insulating state which preserves both the U(1) and
the translational symmetries of the Hamiltonian. In order to see the
effects of quantum fluctuations on the stability of the ground state
phases, we have obtained the diagonal and off diagonal order
parameters using the cluster mean field theory with larger clusters.
Furthermore, using linear spin wave theory we have studied the
behaviors of spin wave excitations and the amount of quantum
fluctuations around the mean field ground state to see the stability
of the ground state phases of the model. We have demonstrated that
in contrast with the standard Bose Hubbard model in which
dipole-dipole interactions or long range hoppings are necessary to have an stable checkerboard supersolid phase, our inhomogeneous bosonic mixture with nearest and next nearest neighbor interaction possesses the
checkerboard supersolid phase. This stability is attributed to the
difference in the nilpotency conditions between $a$ and $b$ bosons.

We have also studied the behavior of the excitation energies in each
phase to get more insights on the ground state phases of the model.
The excitation modes of the solid phases are gapped and the lowest
energy mode has quadratic dispersion around the ordering vectors
$\kv=(0,0)$. All the supersolids and also the superfluid, however,
possess a gapless Goldstone mode with linear dispersion around this
ordering vector. The appearance of this zero mode is a result of the
$U(1)$ symmetry breaking due to the superfluidity long range order.

We have also investigated the behavior of the excitation energies
around all phase transition points to figure out the reason of first
and second order phase transitions. We found out that the abrupt and
smooth changes in the behavior of the low energy excitation modes
and the roton gap responsible of all first and second order phase
transitions, respectively. For example, softening of the linear
gapless mode which is accompanied by vanishing of the roton minimum
at $\kv=(0,0)$, results the second order supersolid-solid phase
transitions.

We have finally investigated the effects of thermal fluctuations on
the stability of the ground state phases at finite temperature. We
have found that at non-zero temperature, in addition to the ground
state phase, other various phases also emerge in the temperature
phase diagram. We have shown that the checkerboard supersolid order
can persist at finite temperatures comparable with the interaction
energies of bosons.

Study of the thermodynamic and magnetocaloric properties of our IBM, and also quantum phase diagram of the mixture on a square
lattice with added intra-component interaction are left for the
future works.

\begin{acknowledgments}
The authors would like to thank Stefan Wessel for insightful
comments on the manuscript. We also thank Marcello Dalmonte for
reading the manuscript and introducing some references. Useful
discussions with Rosario Fazio, Alexander Nersesyan, Sebastiano
Pilati and Sandro Sorella are acknowledged. JA also thanks ICTP
where the initial form of this paper for submission was prepared.
\end{acknowledgments}
\appendix
\section{Origin of the IBM Hamiltonian}\label{appendixa}

In this appendix we will explain how the IBM Hamiltonian introduced
in Eq. (\ref{BH1}), originates from an standard two-orbital
Bose-Hubbard model\cite{Batista2004}. Let us consider a bipartite
model of hard core bosons including two kinds of bosons: a single
orbital boson $a$ and a two-orbital boson $b_{\alpha}$
($\alpha=\{1,2\}$), interacting via the Hamiltonian:
\begin{eqnarray}
\no H_{2B} &=& -t\sum_{\la i,j\ra, \alpha}(\dg a_i b_{\alpha j}+ h.c.)+ U\sum_i n_{1i} n_{2i}\\
\no&+& V_1\sum_{\la i,j\ra} n^a_i (n_j-1) \\
\no&+& V_2\sum_{\la\la i,j\ra\ra}[n^a_i n^a_j + (n_i-1)(n_j-1)] \\
&-&\sum_i \left(\mu^a n^a_i + \mu^b n_i\right), \label{2FBH}
\end{eqnarray}
where, $n^a_i = \dg a_i a_i$ and $n_i=n_{1i}+n_{2i}$ with $n_{\alpha
i} = \dg b_{\alpha i}b_{\alpha i}$. The local Hilbert space of this
bipartite system is a product of the local Hilbert spaces of the
subsystems I and II. The dimension of the local Hilbert space of the
subsystem I is $D_I=2$, since per lattice site $i$ we can only have
the states $\{ \ket{0}_I, \dg a_i \ket{0}_I \}$ where, $\ket{0}_I$
is the vacuum state in subsystem I. The dimension of the local
Hilbert space of the subsystem II is $D_{II}=4$, since the states
per lattice site $j$ are $\{ \ket{0,0}_{II}, \dg b_{1j}
\ket{0,0}_{II}, \dg b_{2j} \ket{0,0}_{II}, \dg b_{1j} \dg b_{2j}
\ket{0,0}_{II}\}$ where $\ket{0,0}_{II}$ is the vacuum state in
subsystem II.

By defining the following new operators:
\begin{equation}
\dg {\tilde b}_j=\frac{1}{\sqrt{2}} (\dg b_{1j} - \dg b_{2j}),~~~
\dg b_j=\frac{1}{\sqrt{2}} (\dg b_{1j} + \dg b_{2j}), \label{defSA}
\end{equation}
and applying them to the vacuum state $\ket{0,0}_{II}$ one can get
the following antisymmetric and symmetric states:
\begin{eqnarray}
\no \ket{\tilde{\Psi}_j} &=& \dg{\tilde{b}_j} \ket{0,0}_{II}=\frac{1}{\sqrt{2}}(\ket{1,0}-\ket{0,1}),\\
\no\ket{\Psi_j} &=& \{ \ket{0,0}_{II}, \dg b_j \ket{0,0}_{II}, \dg b_j \dg b_j \ket{0,0}_{II} \}\\
&=& \{\ket{0,0},\frac{1}{\sqrt{2}}(\ket{1,0}+\ket{0,1}),\ket{1,1}\}.
\label{ASstate}
\end{eqnarray}
These states generate another basis for the local Hilbert space of
the subsystem II with $D_{II}=4$ $(n_j=\dg
{\tilde{b}}_j\tilde{b}_j+\dg b_jb_j)$. By writing the Hamiltonian
(\ref{2FBH}) in terms of the above symmetric and antisymmetric
operators, the Hamiltonian decouples into two symmetric and
antisymmetric parts, which would be a very helpful step for studying
the ground state properties of the model. Before doing this step,
some remarks on the local algebra of these operators are in order.
Following we will show that the local algebra satisfied by
$\dg{\tilde{b}_j}$ and $\dg b_j$ is not the same as the one
satisfied by $\dg{b_{1j}}$ and $\dg{b_{2j}}$.

Let us label the symmetric states as $\{ \ket{0}, \ket{1}, \ket{2}
\}\equiv\{ \ket{0,0}_{II}, \dg b \ket{0,0}_{II}, \dg b \dg b
\ket{0,0}_{II} \}$. Action of the operators $\dg b$, $b$ and $n^b$
on these states leads to the following relations:
\begin{equation}
\begin{split}
\dg b \ket{0} &=\ket{1},~~~ b \ket{0} = 0,~~~~~ n^b \ket{0} = 0,\\
\dg b \ket{1} &=\ket{2},~~~ b \ket{1} = \ket{0},~~~ n^b \ket{1} = \ket{1},\\
\dg b \ket{2} &=0,~~~~ b \ket{2} = \ket{1},~~~~~ n^b \ket{2} = 2
\ket{2}.
\end{split}
\end{equation}
From the above relations one can conclude that the states $ \ket{0},
\ket{1}$ and $\ket{2}$, respectively contain 0, 1 and 2 numbers of
$b$ particles, and $n^b$ is the number operator of $b$-particles.
Moreover the action of the antisymmetric operators $\dg{\tilde{b}}$,
$\tilde{b}$ on the antisymmetric state is given by
\begin{equation}
\begin{split}
\dg {\tilde b} \ket{\tilde{\Psi}} &=- \ket{2},~~~ {\tilde b}
\ket{\tilde{\Psi}} = \ket{0}.
\end{split}
\end{equation}
Writing down the operators $\dg b$, $b$, $n^b$, $\dg{\tilde{b}}$,
and $\tilde{b}$ in terms of the symmetric and antisymmetric states
as:
\begin{equation}
\begin{split}
\dg b &= \ket{1} \bra{0} + \ket{2} \bra{1},~~~ b = \ket{0} \bra{1} + \ket{1} \bra{2}, \\
n^b &= \ket{\tilde{\Psi}} \bra{\tilde{\Psi}} + \ket{1} \bra{1} + 2
\ket{2} \bra{2}, \label{symm}
\end{split}
\end{equation}
and
\begin{equation}
\begin{split}
\dg {\tilde b} &= \ket{\tilde{\Psi}} \bra{0} - \ket{2}
\bra{\tilde{\Psi}},~~~ \tilde b = \ket{0} \bra{\tilde{\Psi}} -
\ket{\tilde{\Psi}} \bra{2},
\end{split} \label{anti}
\end{equation}
we can obtain the following unusual commutation relations:
\begin{equation}
\begin{split}
\no [b, b] &= [\dg b, \dg b]= 0,\\
\no [b, \dg b] &= \ket{0} \bra{0} - \ket{2} \bra{2} ={\mathbf 1}-n^b,\\
[n^b, \dg b] &= \ket{1} \bra{0} + \ket{2} \bra{1} =\dg b,
\end{split}
\end{equation}
where ${\bf 1}$ is the identity matrix for the subsystem II which is
written as ${\bf 1}=\ket{0} \bra{0} + \ket{1} \bra{1} + \ket{2}
\bra{2} + \ket{\tilde{\Psi}} \bra{\tilde{\Psi}}$. Also in the
subsystem I we have:
\begin{equation}
\begin{split}
\dg a &= \ket{1}_I \bra{0}_I,~~~ a = \ket{0}_I \bra{1}_I,~~~ n^a =
\ket{1}_I \bra{1}_I
\end{split} \label{a-boson}
\end{equation}
Now let us go back to the Hamiltonian (\ref{2FBH}). Substituting the
transformations (\ref{defSA}) into the Hamiltonian (\ref{2FBH}), and
then using the relations (\ref{symm}), (\ref{anti}), and
(\ref{a-boson}), the Hamiltonian is transformed to (apart from a
constant):
\begin{eqnarray}
\no H &=& -\sqrt{2} t\sum_{\la i,j\ra}(\dg a_i b_{j} + h.c.) + \frac{U}{2} \sum_i n^{b}_i (n^{b}_i -1) \\
\no &+& V_1\sum_{\la i,j\ra} n^a_i (n^{b}_j -1) \\
\no &+& V_2\sum_{\la\la i,j\ra\ra} (n^{a}_i n^{a}_j + (n^{b}_i-1) (n^{b}_j-1))\\
\label{SBH} &-& \sum_i (\mu^a n^a_i + \mu^b n^b_i).
\end{eqnarray}
Using the following rescalings
\begin{eqnarray}
& \sqrt{2} t \to t,~~ \frac{U}{2} \to U, ~~ V_1 \to V_1,~~  V_2 \to V_2, \\
& \no \mu^a \to \mu^a + 4 V_1,~~ \mu^b \to \mu^b + 2 U + 8 V_2
\end{eqnarray}
the Hamiltonian (\ref{SBH}) is simplified to:
\begin{eqnarray}
\no H &=& -t\sum_{\la i,j\ra}(\dg a_i b_{j}+h.c.)+U\sum_i n^b_i n^b_i\\
\no &+& V_1\sum_{\la i,j\ra}n^a_i n^b_j
+V_2 \sum_{\la\la i,j\ra\ra}(n^a_i n^a_j+n^b_i n^b_j)\\
\label{SBH1} &-& \sum_i (\mu^a n^a_i + \mu^b n^b_i),
\end{eqnarray}
which is the Hamiltonian we have introduced in Eq. (1) of the
manuscript. As it is clearly seen the resulted Hamiltonian is
obtained in terms of the symmetric operators $\dg b$ and $b$, which
means that the local Hilbert space dimension of the subsystem II is
effectively $D_s=3$. As the dimension of the symmetric space is 3 we
have employed a simple boson-spin transformations and mapped this
subsystem to a system of spin one.


\section{Generalized boson-spin transformations for $b$ bosons}\label{appendixb}

In this appendix we explain briefly how the uncommon nilpotency
condition of the $b$ bosons and the fractional exclusion statistics
in Eq. \ref{commutation-b} lead to the boson-spin mapping of $b$
bosons in Eq. \ref{BO}. From the commutation relations in Eq.
\ref{commutation-b}, we find that: 1) the number operator of $b$
bosons possesses the relation $\dg {(\hat{n}^b_i)}=\hat{n}^b_i$ and
2) the number operator of $b$ bosons is not equal to $\dg b b$, i.e.
$\hat{n}^b_i\neq \dg b_ib_i$. Actually, due to the non-canonical
statistics this kind of bosons are different from the canonical one,
and the action of the creation(annihilation) operator $\dg b(b)$ on
the state $\ket{n_b}$ does not lead to
$\sqrt{n_b}\ket{n_b+1}(\sqrt{n_b}\ket{n_b-1})$.

As the local space of the $b$ particle is isomorphic with the Hilbert space of a spin 1, employing the following correspondence of boson-spin basis
\begin{equation}
| 0 \ra \to | 1,-1 \ra,~ | 1 \ra \to | 1,0 \ra,~ | 2 \ra \to |
1,+1 \ra,
\end{equation}
and
\begin{equation}
\tau^+\ket{1,-1}=\sqrt{2}\ket{1,0},~~~ \tau^+\ket{1,0}=\sqrt{2}\ket{1,2},~~~\tau^+\ket{1,2}=0,
\end{equation}
we find the following relations:
\begin{equation}
\t^z=n^b-1, ~~~ \t^+=\alpha \dg b,
\end{equation}
where the coefficient $\alpha$ is readily obtained as follows:
\begin{equation}
\begin{split}
&\tau^+ | 1,-1 \ra = \sqrt{2} | 1,0 \ra,\\
&\to \t^+=\sqrt{2} \dg b, ~~~\t^-=\sqrt{2} b.
\end{split}
\end{equation}
These kinds of mapping between fractional hard core bosons and spin operators are usually employed for different standard and non-standard boson Hamiltonian which could be seen in Ref. [\onlinecite{Batista2004}]


\section{Diagonalization of the spin wave Hamiltonian}\label{appendix}


In order to diagonalize the bilinear part of the spin wave
Hamiltonian $H'$ in Eq. (\ref{SWH}), the first step in an standard
approach, is definition of a Fourier transformation for boson
operators $\hat{a}_i$ and $\hat{b}_i$.
Before going to this step, we notice that  since the phases: MI(4/6), SF and Full preserve the translational symmetry of the original Hamiltonian the classical background has a two-sublattice structure and the excitations of these phases are achieved by defining the two HP bosons: $\hat{a}$ and $\hat{b}$.
However, in other solid and supersolid phases, according to the translational symmetry breaking, the classical background has a four-sublattice structure and more HP bosons should be employed to attain the excitation spectra of these phases.
In this respect, we consider a general background and divide the subsystem with spin $\s$ to two sublattices with bosons $\hat{a}$ and $\hat{c}$, and the subsystem with spin $\t$ to two sublattices with bosons $\hat{b}$ and $\hat{d}$.
Defining the primitive vectors as in the top-center of Fig. \ref{fig:SW-Excitations} and utilizing the following Fourier transformations;
\begin{equation}
\begin{split}
\nonumber \hat{a}_j&=\frac{1}{\sqrt{N/2}} \sum_{\kv} e^{-i \kv\cdot\textbf{r}_j} \hat{a}_{\kv}, ~~ \hat{c}_j=\frac{1}{\sqrt{N/2}} \sum_{\kv} e^{-i \kv\cdot\textbf{r}_j} \hat{c}_{\kv}, \\
\nonumber \hat{b}_j&=\frac{1}{\sqrt{N/2}} \sum_{\kv} e^{-i
\kv\cdot\textbf{r}_j} \hat{b}_{\kv}, ~~ \hat{d}_j=\frac{1}{\sqrt{N/2}}
\sum_{\kv} e^{-i \kv\cdot\textbf{r}_j} \hat{d}_{\kv},
\end{split}
\end{equation}
where $N/2$ is the number of each HP boson, the bilinear Hamiltonian is readily obtained as:
\begin{eqnarray}
\nonumber H^{\prime}=\sum_{\kv} \psi_\kv^\dag H_\kv \psi_\kv,\label{K-Hamiltonian}
\end{eqnarray}
with
$\psi_\kv$, the following 8-component vector:
\begin{eqnarray}
\nonumber \psi_\kv^\dag &=&
\begin{pmatrix}
\nonumber \hat{a}_\kv^{\dag} & \hat{b}_\kv^{\dag} &
\hat{c}_\kv^{\dag} & \hat{d}_\kv^{\dag} & \hat{a}_{-\kv} &
\hat{b}_{-\kv} & \hat{c}_{-\kv} & \hat{d}_{-\kv}
\end{pmatrix},
\end{eqnarray}
and
\begin{eqnarray} \label{eq:HK}
H_\kv =
\begin{pmatrix}
A & B \\ B^* & A^*
\end{pmatrix},
\end{eqnarray}
where $A$ and $B$ are two $4$-square matrices with complex functions.
The general forms of the matrices $A$ and $B$ are given by
\begin{eqnarray}
\nonumber A &=&
\nonumber \begin{pmatrix}
\nonumber \alpha_{11} & \alpha_2^{\ast} & \alpha_9 & \alpha_6^{\ast} \\
\nonumber \alpha_2 & \alpha_{12} & \alpha_8 & \alpha_{10} \\
\nonumber \alpha_9 & \alpha_8^{\ast} & \alpha_{13} & \alpha_4^{\ast} \\
\nonumber \alpha_6 & \alpha_{10} & \alpha_4 & \alpha_{14}
\end{pmatrix}, ~
\nonumber B =
\nonumber \begin{pmatrix}
\nonumber 0 &  \alpha_1^{\ast} & \alpha_9 & \alpha_5^{\ast} \\
\nonumber \alpha_1^{\ast} & 0 & \alpha_7^{\ast} & \alpha_{10} \\
\nonumber \alpha_9 & \alpha_7^{\ast} & 0 & \alpha_3^{\ast} \\
\nonumber \alpha_5^{\ast} & \alpha_{10} & \alpha_3^{\ast} & 0
\end{pmatrix},
\end{eqnarray}
where
\begin{equation}
\begin{split}
\alpha_1 & = 2 w^{11}_{ab}\cos{(k_x a'/2)},~~
\alpha_2=2 w^{12}_{ab}\cos{(k_x a'/2)},\\
\alpha_3&=2 w^{11}_{cd}\cos{(k_x a'/2)},~~
\alpha_4 = 2 w^{12}_{cd}\cos{(k_x a'/2)}, \\
\alpha_5 & =2 w^{11}_{ad}\cos{(k_y a'/2)},~~
\alpha_6  =2 w^{12}_{ad}\cos{(k_y a'/2)}, \\
\alpha_7 & =2 w^{11}_{cb}\cos{(k_y a'/2)},~~
\alpha_8 =  2 w^{12}_{cb}\cos{(k_y a'/2)}, \\
\alpha_9 & =4 V_2 g^1_{ac} \cos{(k_x a'/2)} \cos{(k_y a'/2)}, \\
\alpha_{10} & =4 V_2 g^1_{bd} \cos{(k_x a'/2)} \cos{(k_y a'/2)}, \\
\alpha_{11} & =2(w^{23}_{ab}+w^{23}_{ad})+4 V_2 g^2_{ac} - h e_a, \\
\alpha_{12} & =2(w^{34}_{ab}+w^{34}_{cb})+4 V_2 g^2_{bd} - h e_b, \\
\alpha_{13} & =2(w^{23}_{cb}+w^{23}_{cd})+4 V_2 g^3_{ac} - h e_c, \\
\label{alpha-4} \alpha_{14} & =2(w^{34}_{ad}+w^{34}_{cd})+4 V_2 g^3_{bd} - h e_d,
\end{split}
\end{equation}
with
\begin{equation}
w^{\alpha\beta}_{mn}=V_1 g_{mn}^\alpha-J f_{mn}^\beta.\label{w}
\end{equation}
Here, $a'$ is the length of the primitive vectors shown in the top-center of Fig. \ref{fig:SW-Excitations}, $\alpha$ and $\beta$ are 1, 2, 3, 4 and $m$ and $n$ are the sublattices label: $a, b, c$ and $d$.
The coefficients $f_{mn}^\beta$, $g_{mn}^\alpha$ and $e_m$ are given in terms of $\theta_m, \theta_n$ and $\phi_m, \phi_n$ as follows:
\begin{equation}
\begin{split}
f^1_{mn} & =\sqrt{S_m S_n} ((\cos{\theta_m} \cos{\theta_n} -1) \cos{(\phi_m - \phi_n)}, \\
&+i \sin{(\phi_m - \phi_n)} (\cos{\theta_n} - \cos{\theta_m})), \\
f^2_{mn} & =\sqrt{S_m S_n} ((\cos{\theta_m} \cos{\theta_n} +1) \cos{(\phi_m - \phi_n)}, \\
&+i\sin{(\phi_m - \phi_n)} (\cos{\theta_n} + \cos{\theta_m})), \\
f^3_{mn} & =-2 S_n \sin{\theta_m} \sin{\theta_n} \cos{(\phi_m - \phi_n)}, \\
f^4_{mn} & =-2 S_m \sin{\theta_m} \sin{\theta_n} \cos{(\phi_m - \phi_n)}, \\
g^1_{mn}&=\frac 12\sqrt{S_m S_n} \sin{\theta_m} \sin{\theta_n}, \\
g^2_{mn} & =-S_n \cos{\theta_m} \cos{\theta_n}, \\
g^3_{mn} & =-S_m \cos{\theta_m} \cos{\theta_n}, \\
e_{m} & =- \cos{\theta_m},
\end{split}
\end{equation}
where $S_m$ and $S_n$ are the spins of sublattices $m$ and $n$, respectively. Performing a paraunitary transformation $T$, the Hamiltonian $H_\kv$ in Eq. (\ref{eq:HK}) is diagonalized as:
\begin{eqnarray}
\nonumber\psi_\kv^{\dag} H_\kv \psi_\kv = \psi_\kv^{\dag} T^{\dag} (T^{\dag})^{-1} H_\kv T^{-1} T \psi_\kv = \Psi^{\dag}_\kv \mathcal{E}_\kv \Psi_\kv,
\end{eqnarray}
where $\mathcal{E}_\kv$ is the para-diagonalized matrix containing the excitation energies and $\Psi_\kv=T\psi_\kv$ is a para-vector of new bosonic operators.
The paraunitary transformation satisfies the following relations
\begin{eqnarray}
T \hat I T^{\dag} = \hat{I}, ~~~~ T^{\dag} \hat{I} T = \hat I, ~~~~ T^{\dag} \hat{I} = \hat{I} T^{-1},
\end{eqnarray}
with
\begin{eqnarray}
\hat{I}_{8 \times 8} =
\begin{pmatrix}
\nonumber I_{4\times 4} & 0 \\
\nonumber 0 & -I_{4\times 4} \\
\end{pmatrix},
\end{eqnarray}
where $I_{4\times 4}$ is the $4\times 4$ unitary matrix. In order to
obtain the paraunitary transformation $T$ we utilize the
following procedure which introduced by Colpa for a
positive-definite Hamiltonian\cite{colpa1978diagonalization}. First
we write the Hamiltonian $H_\kv$ as $H_\kv = \kappa ^{\dag}_\kv
\kappa_\kv$ where the matrix $\kappa_\kv$ is the Cholesky
decomposition of the Hamiltonian. Then, we find the unitary
transformation matrix, $\upsilon_\kv$ which diagonalizes the
hermitian matrix $\kappa_\kv\hat{I}\kappa^{\dag}_\kv$ as
$\mathcal{L}_\kv=
\upsilon^{\dag}_\kv(\kappa_\kv\hat{I}\kappa^{\dag}_\kv)\upsilon_\kv$.
The diagonal matrix $\mathcal{E}_\kv$ is readily obtained from the
relation $\mathcal{E}_\kv= \hat{I}\mathcal{L}_\kv$. Finally, by
solving the equation $\upsilon_\kv\sqrt{\mathcal{E}_\kv}=\kappa_\kv
T^{-1}$ row to row, we achieve the paraunitary transformation $T$.

The above diagonalization procedure is for the general case of
four-sublattice structure which is employed for the $a$CS(5/6), $b$CS(4/6) and CS(3/6) solid, and for $a$CSS, $b$CSS and CSS supersolid phases.
For the MI(4/6), SF and Full phases where the translational symmetry in both subsystems is preserved, the MF ground states are given by a two-sublattice structure and the $4\times 4$ matrices $A$ and $B$ are simplified to two $2\times 2$ matrices.
Following we will obtain these matrices. Using the Fourier transformations;
\begin{eqnarray}
\nonumber \hat{a}_j=\frac{1}{\sqrt{N}} \sum_{\kv} e^{-i \kv\cdot\mathbf{r}_j} \hat{a}_{\kv}, ~~ \hat{b}_j=\frac{1}{\sqrt{N}} \sum_{\kv} e^{-i
\kv\cdot\mathbf{r}_j} \hat{b}_{\kv},
\end{eqnarray}
where $N$ is the number of each HP boson and $\mathbf{r}_j$ is given in terms of the primitive vectors shown in the top-left of Fig. \ref{fig:SW-Excitations}, the bilinear Hamiltonian is readily obtained as:
\begin{eqnarray}
\nonumber H^{\prime}=\sum_{\kv} \psi_\kv^\dag H_\kv \psi_\kv,\label{K-Hamiltonian}
\end{eqnarray}
with
\begin{eqnarray}
\nonumber \psi_\kv^\dag &=&
\begin{pmatrix}
\nonumber \hat{a}_\kv^{\dag} & \hat{b}_\kv^{\dag} & \hat{a}_{-\kv} &
\hat{b}_{-\kv}
\end{pmatrix},
\end{eqnarray}
and
\begin{eqnarray} \label{eq:HK2}
H_\kv =
\begin{pmatrix}
A & B \\ B^* & A^*
\end{pmatrix},
\end{eqnarray}
where $A$ and $B$ are given by
\begin{eqnarray}
\nonumber A &=&
\nonumber \begin{pmatrix}
\nonumber \alpha_{3} & \alpha_2^{\ast} \\
\nonumber \alpha_2 & \alpha_{5} \\
\end{pmatrix}, ~
\nonumber B =
\nonumber \begin{pmatrix}
\nonumber \alpha_4 & \alpha_1^{\ast} \\
\nonumber \alpha_1^{\ast} & \alpha_6 \\
\end{pmatrix},
\end{eqnarray}
where
\begin{equation}
\begin{split}
 \alpha_1 & =  2 w^{11}_{ab}\cos(k_x a/2) \cos(k_y a/2),\\
 \alpha_2 & =  2 w^{12}_{ab}\cos(k_x a/2) \cos(k_y a/2),\\
 \alpha_3 & =  4 w^{23}_{ab}-h e_a\\
& + 2 V_2 \left(g^2_{aa}+g^3_{aa} + g^1_{aa}(\cos(k_x a)+\cos(k_y a)\right),\\
 \alpha_4 & =  V_2 g^1_{aa} (\cos(k_x a) + \cos(k_y a)), \\
 \alpha_5 & =  4 w^{34}_{ab}- h e_b \\
&+ 2 V_2 (g^2_{bb} + g^3_{bb} + g^1_{bb}(\cos(k_x a) + \cos(k_y a)), \\
 \alpha_6 & =  V_2 g^1_{bb} (\cos(k_x a) + \cos(k_y a)).\\
\end{split}
\end{equation}


\bibliography{draft-PRB-07July17}

\end{document}